\documentclass[aps,amssymb,showpacs,floatfix, twocolumn, longbibliography]{revtex4-2}
\usepackage{amssymb,graphicx}
\usepackage{epsfig,graphicx}
\usepackage{subcaption}
\usepackage{subfloat}
\usepackage{epstopdf}
\usepackage{amsmath}
\usepackage{xcolor}
\usepackage{romannum}
\usepackage{tikz}
\usepackage[colorlinks=true,citecolor=blue,linkcolor=blue,hypertexnames=false]{hyperref}

\begin{document}
\pagenumbering{arabic}

\title{Localized states, topology and anomalous Hall conductivity on a 30 degrees twisted bilayer honeycomb lattice}

\author{Grigory Bednik}
\affiliation{Department of Physics, University of Nebraska Omaha}

\date{August 22, 2025}

\begin{center}
\begin{abstract}


We consider $30^{\circ}$ twisted bilayer formed by two copies of Haldane model and explore its evolution with varying interlayer coupling strength. Specifically, we compute the system's energy spectrum, its fractal dimensions, topological entanglement entropy, local Chern markers and anomalous Hall conductivity. We find that at weak interlayer coupling, the system 
still has a bulk energy gap, topological edge states 
and retains topological properties of the isolated layers, but at strong interlayer coupling, this energy gap closes. However, at small values of the Haldane mass $m$, another bulk gap opens.
At strong interlayer coupling, the system possesses multiple states localized at various locations of the lattice, including corner states. We emphasize that these corner states do not originate from the topological edge states at the weak coupling, and their location is not necesarily attributed to the bulk gap. We also compute fractal dimensions and establish that the system at large interlayer coupling is multifractal.  Finally, we establish that topological entanglement entropy and anomalous Hall conductivity can be used to characterize the system's topological properties in the same way as a local Chern marker. Our results suggest that the bulk gap at the strong interlayer coupling has non-topological origin.


\end{abstract}
\end{center}

\maketitle

\section{Introduction}

Quasicrystals are a novel class of materials discovered in 1980-s, whose lattice is non-periodic and yet is ordered according to certain rules. 
One possible way to obtain a quasicrystal is to consider a periodic lattice in higher dimension and project it into a plane with a normal vector incommensurate to the original lattice \cite{RevModPhys.93.045001}. Other specific models of quasicrystals include Fibbonacci lattice \cite{RevModPhys.93.045001}, Penrose tiling, Rauzy tiling \cite{PhysRevB.100.214109}, Aubry-Andr\'e model \cite{aubry1980analyticity} etc. It was established that quasicrystals generally host multifractal electronic states, which are neither extended, nor localized, but instead characterized by non-integer fractal dimension \cite{PhysRevB.107.174205, PhysRevA.108.053305, PhysRevLett.108.220401}. 

It is widely known that periodic crystals can host multiple topological phases. 
On the other hand, topological properties of quasicrystals are still not well-understood. In a recent work \cite{PhysRevLett.109.106402}, it  was suggested that topological quasicrystals can be built from crystals in higher dimensions, and their topology can be deduced from crystal topology. In a few other works, empirical models of topological quasicrystals were proposed  \cite{PhysRevB.100.214109, PhysRevB.91.085125, PhysRevLett.129.056403, Jiahao_Fan:13203}, 
which host gapless  edge states and whose topology was characterized directly using empirical 'local Chern markers' \cite{PhysRevB.84.241106}. However, there is still no systematic understanding of how non-trivial topology in quasicrystals may arise.

Motivated by this, we are interested in studying a quasicrystal obtained by smoothly deforming a crystal with known non-trivial topology. Specifically, we choose to consider 30 degrees twisted bilayer honeycomb lattice - a model which is currently actively studied in the context of twisted bilayer graphene \cite{PhysRevB.99.165430, Yu2019, PhysRevB.107.014501, zhang2024criticalfilamentssuperconductivityquasiperiodic}. However we consider a 
 twisted bilayer, in which each monolayer is described by gapped Haldane model \cite{PhysRevResearch.2.033071, PhysRevB.106.125428, sym14081736}.  We explore our model's evolution once the interlayer coupling is turned on. We obtain that at weak coupling, the model retains all properties of Haldane model: its bulk spectrum remains gapped and the topological edge states persist. Thus we show that our system of two coupled copies of Chern insulators can be viewed as a topological quasicrystal. On the other hand, as the interlayer coupling becomes larger, the bulk energy gap closes, and the topological edge states disappear. Yet, since this system is non-periodic, its eigenstates are generally not extended. We obtain numerically that our bilayer possesses multiple kinds of states localized at various locations of the lattice. These include, but not limited to corner states, which were previously found in Refs. \cite{PhysRevB.106.125428, sym14081736}. We also observe that at large interlayer coupling, a new energy gap may open, 
 but we argue that it is not topological. 

In general, characterizing topological properties of a quasicrystal is challenging because most well-established topological concepts (e.g. Chern numbers) were developed in momentum space, i.e. only for crystalline systems. Nevertheless, we are still able to find out if the system is topological by computing its topological entanglement entropy \cite{PhysRevLett.96.110404} (we use computational methods previously considered in \cite{Helmes:2016ahb, Ingo_Peschel_2003, PhysRevB.94.125142, PhysRevB.99.155153}). Furthermore, we study its topological properties by computing 'local Chern marker' \cite{PhysRevB.84.241106}. Finally we numerically compute its anomalous Hall conductivity by using Kubo formula and find that it can be used to characterize topological properties of the quasicrystal in the same way as local Chern marker. 

This paper is organized as follows. In Sec. \ref{Sec:Model} we introduce our model, and in Sec. \ref{Sec:Phase_diagram} we describe its phase diagram, which consists of multiple phases: topological weakly-coupled, non-topological weakly coupled and strongly coupled multifractal. The latter may have both gapped and gapless phases. We also comment on multifractal properties of our model (Sec. \ref{Sec:Multifractal_properties}).  In Sec. \ref{Sec:Topological properties} we describe its topological properties. Specifically, in Sec.  \ref{Sec:Topological_entanglement_entropy} we present our results for its topological entanglement entropy and in Sec. \ref{Sec:Local_Chern_marker} we describe its topological properties using the 'local Chern markers'. In Sec. \ref{Sec:Anomalous_Hall_conductivity} we discuss anomalous Hall conductivity. We summarize our findings in Sec. \ref{Sec:Discussion} .


\section{Model}

\label{Sec:Model}

\begin{figure}
	\begin{subfigure}[t]{0.15\textwidth}
	\includegraphics[width=3cm, height=3cm]{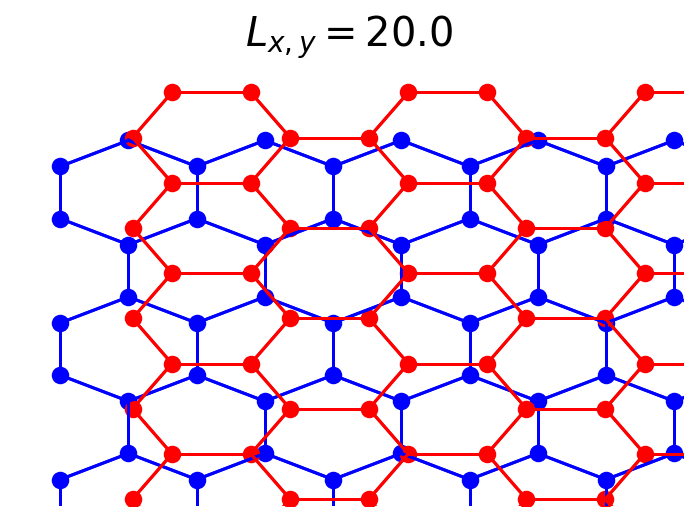}
		\subcaption{}	
		\label{Lattice_Lx_20}
	\end{subfigure}
	\begin{subfigure}[t]{0.15\textwidth}
	\includegraphics[width=3cm, height=3cm]{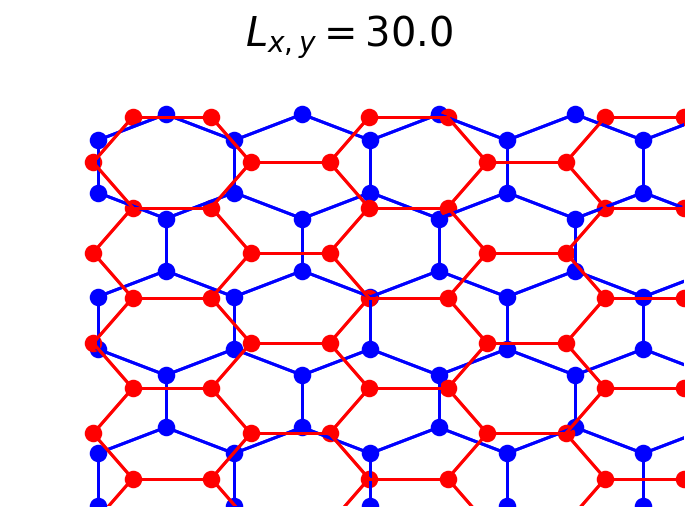}
		\subcaption{}	
		\label{Lattice_Lx_30}
	\end{subfigure}
	\begin{subfigure}[t]{0.15\textwidth}
	\includegraphics[width=3cm, height=3cm]{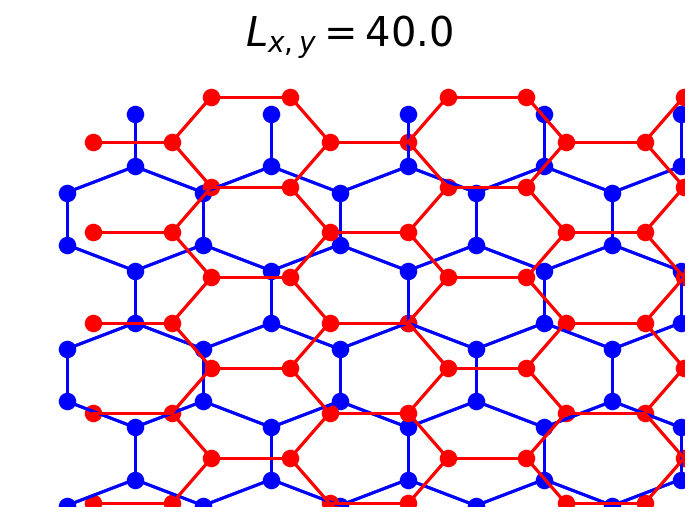}
		\subcaption{}	
		\label{Lattice_Lx_40}
	\end{subfigure}
\caption{A schematic picture of the $30^{\circ}$ twisted bilayer of two honeycomb lattices.
Here we show its top-left corners with $L_{x, y}=20$ (\ref{Lattice_Lx_20}), $L_{x, y}=30$ (\ref{Lattice_Lx_30}), $L_{x, y}=40$ (\ref{Lattice_Lx_40}).
 The top layer (blue) is formed by 
sites with coordinates $\vec{r}_{A, B} = \vec{a}_1 n_1 + \vec{a}_2 n_2 + \vec{c}_{A, B}$, where $\vec{a}_1 = (\sqrt{3}, 0)$ , $\vec{a}_2 = (\sqrt{3}/2, 3/2)$, $\vec{c}_{A, B} = (0, \pm 1)$. The bottom layer (red) is obtained by rotating the top layer at 30 degrees. We consider a system with open boundary conditions, whose sites are located inside a square  $-L_{x, y} < x,y  < L_{x, y}$. The cases (\ref{Lattice_Lx_20}, \ref{Lattice_Lx_30}) correspond to zigzag-armchair edges, and the case (\ref{Lattice_Lx_40}) corresponds to the bearded-armchair edge. }
\label{Lattice}
\end{figure}

\begin{figure}
	\begin{subfigure}[t]{0.2\textwidth}
	\includegraphics[width=3.5cm, height=4cm]{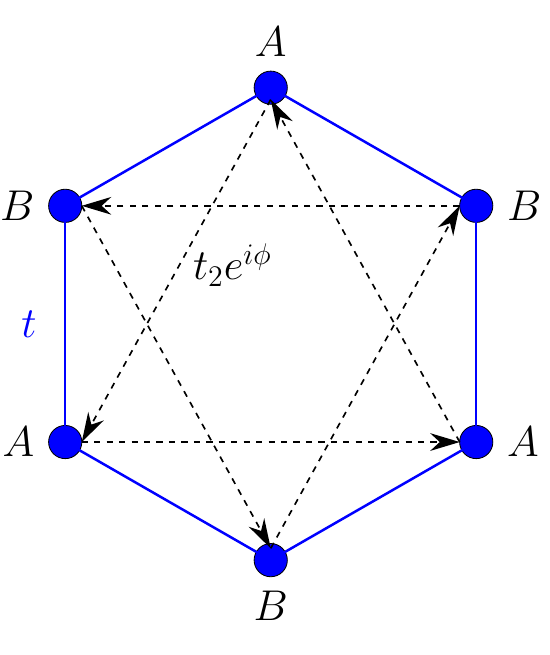}
		\subcaption{}	
		\label{Top_layer_couplings}
	\end{subfigure}
	\begin{subfigure}[t]{0.2\textwidth}
	\includegraphics[width=4.5cm, height=4cm]{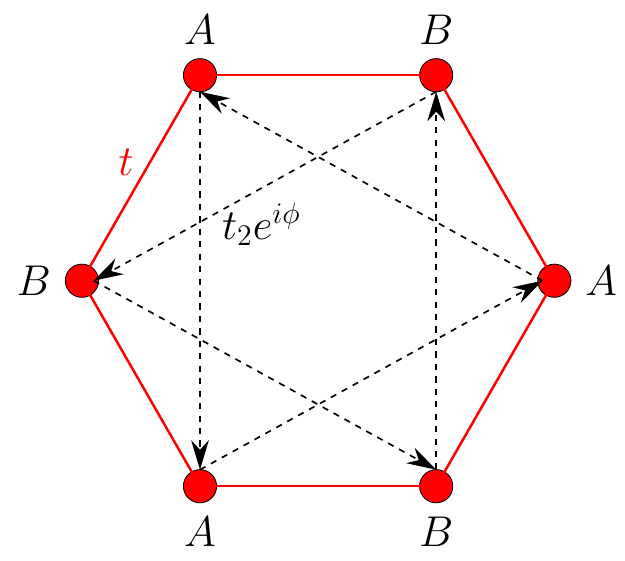}
		\subcaption{}	
		\label{Bot_layer_couplings}
	\end{subfigure}
\caption{A schematic picture of couplings in Haldane model in the top (\ref{Top_layer_couplings}) and bottom (\ref{Bot_layer_couplings}) layers. In each layer, there is a nearest neighbors coupling $t$ between $A$, $B$ sites and a next nearest neighbors coupling $t_2 e^{\pm i \phi}$. The arrows mark the sign of each coupling, namely an arrow pointing from the site $i$ to the site $j$ means that the corresponding term in the Hamiltonian has the form $ t_2 e^{+ i \phi} c_i^{\dagger} c_j $.  The bottom layer can be obtained by rotating the top layer by $30^{\circ}$ degrees.}
\label{Couplings}
\end{figure}

We consider a tight-binding model of electrons on a  bilayer of two honeycomb lattices twisted relative to each other by $30^{\circ}$ angle (see Fig. \ref{Lattice}). Each of the lattices is described by Haldane Hamiltonian, and in addition, electrons from the different layers interact via exponentially decaying potential. The total Hamiltonian has the form
\begin{eqnarray}
H = \sum\limits_{\alpha=1, 2} H_{\alpha} + V_{1-2},
\label{MainHamiltonian}
\end{eqnarray}
where $\alpha$ is the layer index, $H_{\alpha}$ is Haldane Hamiltonian of each layer, and $V_{1-2}$ describes interaction between the layers. Specifically,
\begin{eqnarray}
H_{\alpha} &=& t_{intra} \sum \limits_{<ij>} c_{i, \alpha}^{\dagger} c_{j, \alpha}  
+ t_2 \sum\limits_{\ll i, j\gg} e^{-i \nu_{ij} \phi} c_{i, \alpha}^{\dagger} c_{j, \alpha} 
\nonumber\\
&& + 3 \sqrt{3}m \sum\limits_i \epsilon_i c_{i, \alpha}^{\dagger} c_{i, \alpha},
\label{H_alpha}
\\
V_{1-2} &=& t_{inter} \sum\limits_{i, j} v_{i,j} c_{i, 1}^{\dagger} c_{j, 2} + h.c.
\label{V_1_2}
\end{eqnarray}
Here $i, j$ numerate the lattice sites within each layer, $< i, j >$ is a sum over the nearest neighbors, $\ll i, j \gg$ is a sum over the next-nearest neighbors. $t_{intra}$ is the nearest neghbors hopping (between A, B sites), $t_2$ is the next-nearest neighbors hopping (between A-A or B-B sites), $\nu_{ij} = \pm 1$ depending on the relative direction between $i$ and $j$ (see Fig. \ref{Couplings}), $\phi = \pi/2$, $m$ is Haldane mass, which we assume to be non-negative, and $\epsilon_i = \pm 1$ for A(B) sites (see e.g. ref. \cite{bernevig2013topological} for more details). The interlayer potential has the form
\begin{eqnarray}
v_{i, j} = 
\left\{
e^{- \frac{|\vec{r}_{i,1} - \vec{r}_{j, 2}| }{r_0}},
\quad |\vec{r}_{i,1} - \vec{r}_{j, 2}| < r_{max}
\atop
0, \qquad\qquad\quad |\vec{r}_{i,1} - \vec{r}_{j, 2}| > r_{max}.
\right.
\end{eqnarray}
Here $\vec{r}_{i,1}, \vec{r}_{j, 2}$ denote the positions of the lattice sites $i, j$ located at the layers $1, 2$,     
the distance between them is $|\vec{r}_{i,1} - \vec{r}_{j, 2}|$, and to simplify the model we neglect interlayer spacing. Also we introduced long-distance cutoff $r_{max}$ to decrease the complexity of our numerical calculations. We fix the numerical values of our parameters as $t_{intra}=1$, $t_2=1$, $\phi=\pi/2$,  $r_0 =1$, $r_{max} = 2$ (we checked that changing $t_2$, $r_{max}$ does not have qualitative impact on the model's properties) and explore its evolution over  $m$, $t_{inter}$. Since our 30 degrees twisted bilayer is a non-periodic system, we consider a system forming a square with equal lengths in both directions $L_x=L_y$ and fix open boundary conditions.

\subsection{Phase diagram}

\label{Sec:Phase_diagram}

To study the properties of our model, we compute its energy spectrum and wavefunctions numerically using exact diagonalization. Further, we characterize their spatial properties by computing fractal dimensions
\begin{eqnarray}
D_q = \frac{\log(P_q)}{(q-1) \log(N)},
\label{Frac_dim}
\end{eqnarray}
where $P_q$ is a participation ratio defined as $P_q = \sum_i (\psi^{\dagger}_i \psi_i )^q$, and $i=1\ldots N$ runs over all  lattice sites. Here we consider a few fixed values of $q$, but unless specified otherwise, we assume $q=2$. In a system of large spatial size, fractal dimension is expected to be close to $1$ for extended bulk states, $1/2$ for edge states and $0$ for localized states (e.g. corner modes). In this section, we use $D_q$ to find the most localized states, which have the smallest $D_q$.

First, let us recall our model's properties in the known case when the interlayer coupling $t_{inter}$ is set to zero. This model is a Chern insulator at $m < 1$ (we assume that both layers have the same signs of Chern numbers). Its bulk states form two bands separated by an energy gap, which in turn, is filled by edge states. At $m=1$, the bulk gap closes, and at $m>1$, it reopens again, and the system becomes a trivial insulator. 

\begin{figure}
	\includegraphics[width=6cm,angle=0] 
	{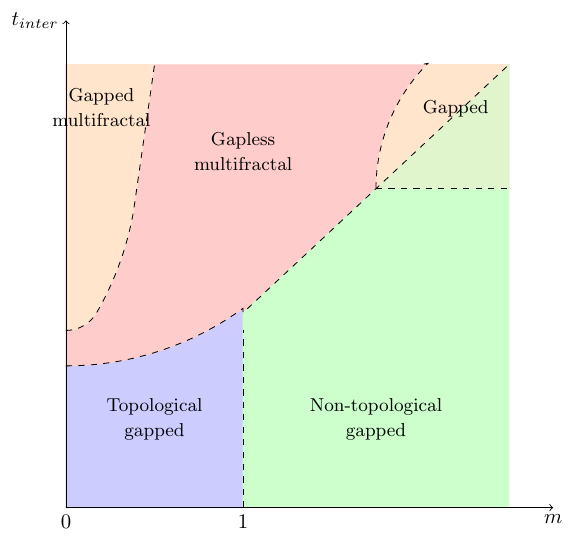}
\caption{Schematic phase diagram of the $30^{\circ}$ degrees twisted bilayer Haldane model defined by the Eq. \ref{MainHamiltonian} obtained by exact diagonalization. At small $t_{inter}$, the model has exactly the same topological/non-topological phases as a single layer Haldane model. At larger $t_{inter}$, the bulk gap closes, and the in-gap edge states disappear. Simultaneously, another bulk gap opens at $m \approx 0$ and $m \gtrsim 1$. }
\label{PhaseDiagram}
\end{figure}

\begin{figure}
	\hspace{-4cm}
	\begin{subfigure}[t]{0.2\textwidth}
		\includegraphics[width=8cm,angle=0] 
		{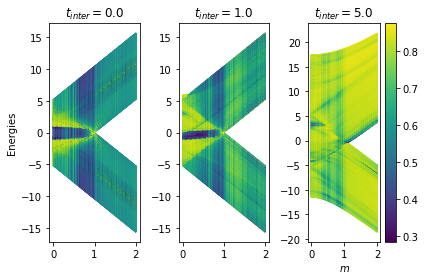}
		\subcaption{}	
		\label{Loop_over_m}
	\end{subfigure}
	\\
	\hspace{-4cm}
\caption{ (\ref{Loop_over_m}) Energies of states plotted against $m$ for various values of $t_{inter}$. We choose $L_x = L_y=30$. The color represents their fractal dimensions  $D_q$ for $q=2$. 
At weak interlayer coupling, the spectrum is gapped for $m>1$, whereas for $m <1$ the gap is filled by the edge states, which have smaller fractal dimensions. At strong inerlayer coupling, the edge states disappear, and the spectrum becomes gapless. 
}
\label{Loop_over_m_and_t_inter}
\end{figure}

\begin{figure}
	\begin{subfigure}[t]{0.15\textwidth}
		\includegraphics[width=3cm,angle=0] {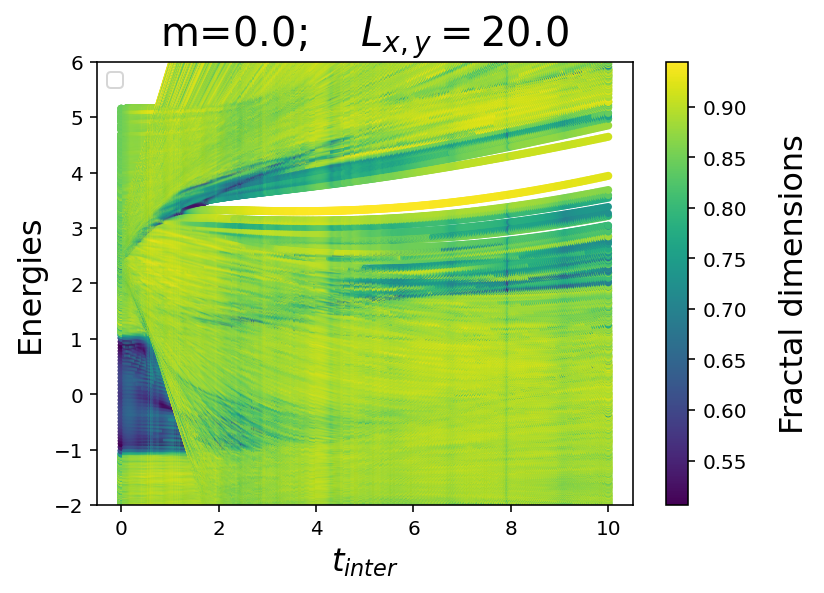}
		\subcaption{}
		\label{Loop_over_t_inter_m_0.0_Lx_20}
	\end{subfigure}
	\begin{subfigure}[t]{0.15\textwidth}
		\includegraphics[width=3cm,angle=0] {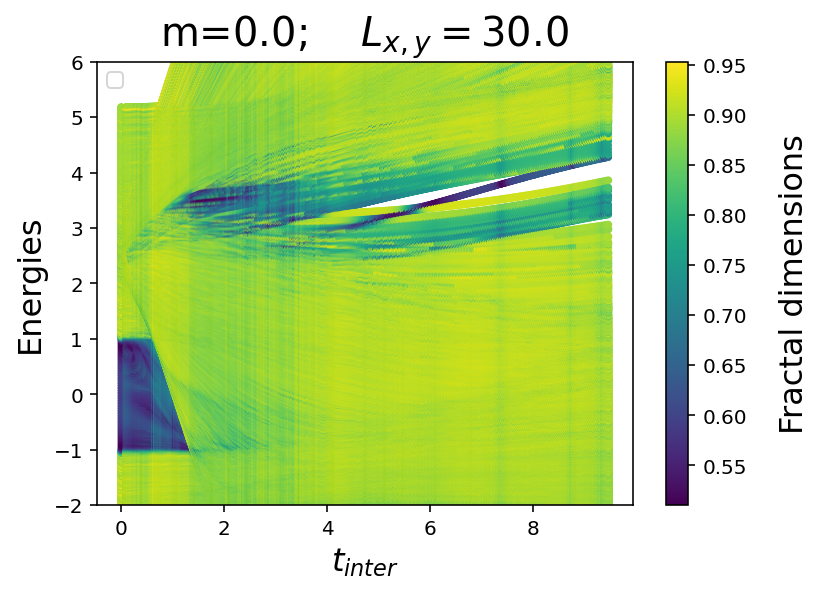}
		\subcaption{}
		\label{Loop_over_t_inter_m_0.0_Lx_30}
	\end{subfigure}
	\begin{subfigure}[t]{0.15\textwidth}
		\includegraphics[width=3cm,angle=0] {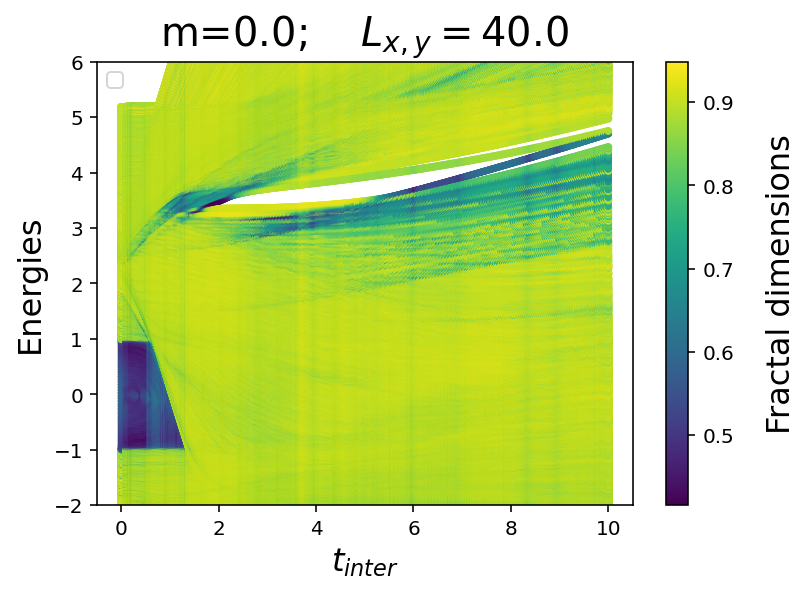}
		\subcaption{}	
		\label{Loop_over_t_inter_m_0.0_Lx_40}
	\end{subfigure}
	\begin{subfigure}[t]{0.15\textwidth}
		\includegraphics[width=3cm,angle=0] {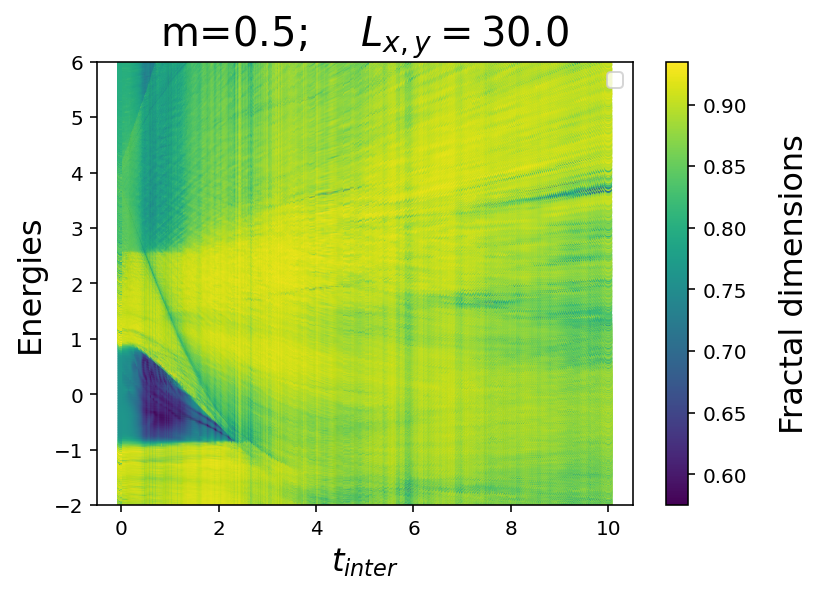}
		\subcaption{}
		\label{Loop_over_t_inter_m_0.5_Lx_30}
	\end{subfigure}
	\begin{subfigure}[t]{0.15\textwidth}
		\includegraphics[width=3cm,angle=0] {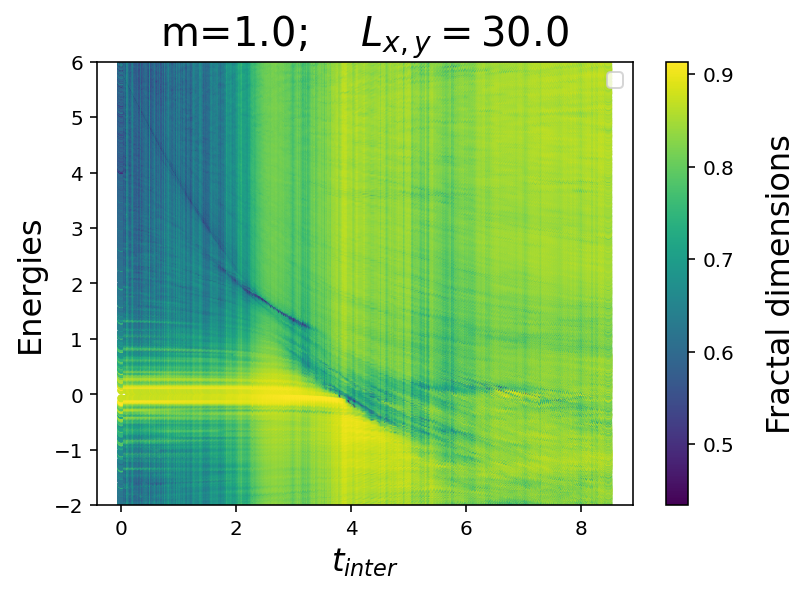}
		\subcaption{}
		\label{Loop_over_t_inter_m_1.0_Lx_30}
	\end{subfigure}
	\begin{subfigure}[t]{0.15\textwidth}
		\includegraphics[width=3cm,angle=0] {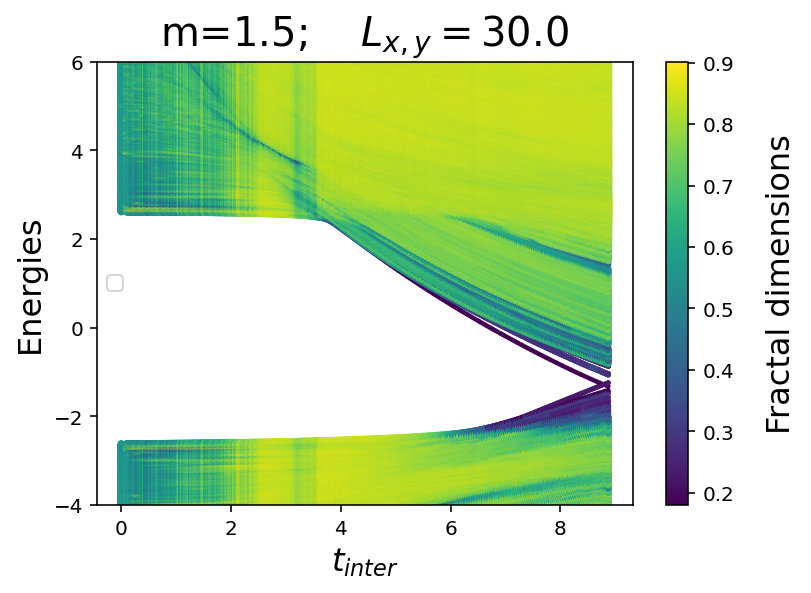}
		\subcaption{}	
		\label{Loop_over_t_inter_m_1.5_Lx_30}
	\end{subfigure}
	\caption{Energies of states plotted against $t_{inter}$ for $m=0$ and $L_{x, y}=20$ (\ref{Loop_over_t_inter_m_0.0_Lx_20}), $L_{x, y}=30$ (\ref{Loop_over_t_inter_m_0.0_Lx_30}), $L_{x, y}=40$ (\ref{Loop_over_t_inter_m_0.0_Lx_40}) as well as   $m=0.5$ (\ref{Loop_over_t_inter_m_0.0_Lx_20}), $m=1.0$ (\ref{Loop_over_t_inter_m_0.0_Lx_20}), $m=1.5$ (\ref{Loop_over_t_inter_m_0.0_Lx_40}) at $L_{x, y}=30$. One can see that at small $t_{inter}$, the system retains properties of its counterpart at $t_{inter}=0$, but at large $t_{inter}$, the system either become gapless ($m=0.5, 1$), or a new gap opens ($m=0, 1.5$). The color represents each state's fractal dimension $D_q$ at $q=2$, and we use the same colorbar as on the Fig. \ref{Loop_over_m_and_t_inter}. Localized states have the smallest fractal dimensions and thus are shown in blue color, whereas more extended states are shown in green.}
	\label{Loop_over_t_inter}
\end{figure}

Once the interlayer coupling  is turned on, the phase diagram of the system in $m ,t_{inter}$ space has multiple phases, which we schematically outline on the Fig. \ref{PhaseDiagram} and which we obtained by running over different values of $m$ and $t_{inter}$ (see Figs. \ref{Loop_over_m_and_t_inter}, \ref{Loop_over_t_inter}). Specifically, if we start from the topological phase of the Haldane model, we observe (see Figs. \ref{Loop_over_m_and_t_inter}) that at small coupling, the bulk states are still separated by an energy gap and the edge states persist, but as $t_{inter}$ is increasing, the bulk energy gap shrinks. Once $t_{inter}$ reaches a critical value ($t_{inter} \approx 1.5$ for $m=0$), the bulk gap closes, and the edge states disappear. As we increase $t_{inter}$ further, at small $m$, another energy gap appears (see Fig. \ref{Loop_over_t_inter_m_0.0_Lx_20}, \ref{Loop_over_t_inter_m_0.0_Lx_30}, \ref{Loop_over_t_inter_m_0.0_Lx_40} ). However, we do not observe it at larger $m$ (see Fig. \ref{Loop_over_t_inter_m_0.5_Lx_30} ). Similarly, if we start from the non-topological phase (see Fig. \ref{Loop_over_t_inter_m_1.5_Lx_30}), at small $t_{inter}$, the system remains gapped, and as $t_{inter}$ is increasing, the gap becomes smaller and eventually disappears. Near this phase transition, another gap may appear.

\begin{figure}
	\begin{subfigure}[t]{0.2\textwidth}
		\hspace{-1cm}
		\includegraphics[width=4cm,angle=0] {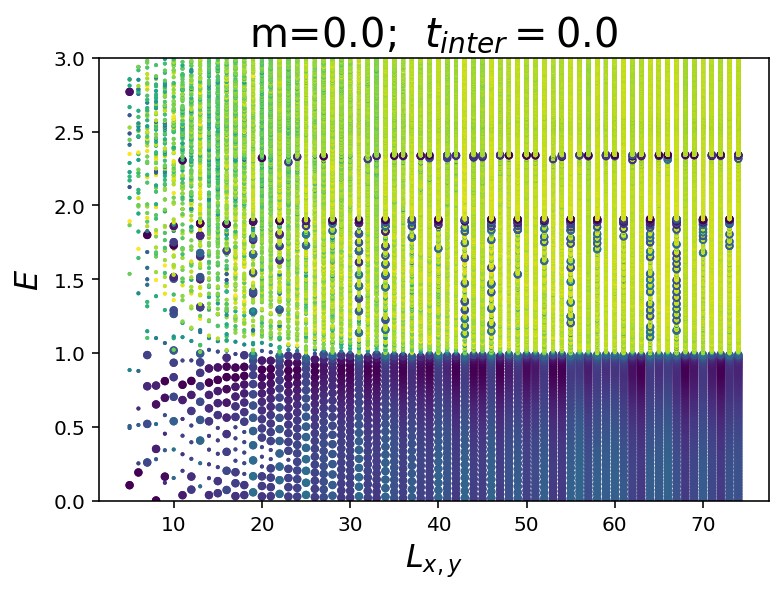}
		\subcaption{}
		\label{Loop_over_Lx_m_0.0_t_inter_0.0}
	\end{subfigure}
	\begin{subfigure}[t]{0.2\textwidth}
		\includegraphics[width=4cm,angle=0] {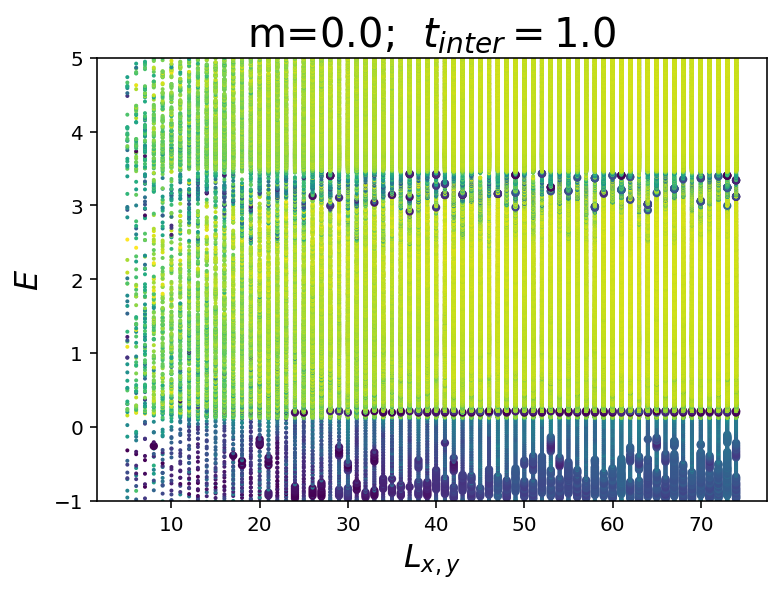}
		\subcaption{}
		\label{Loop_over_Lx_m_0.0_t_inter_1.0}
	\end{subfigure}
	\begin{subfigure}[t]{0.2\textwidth}
		\hspace{-1cm}
		\includegraphics[width=4cm,angle=0] {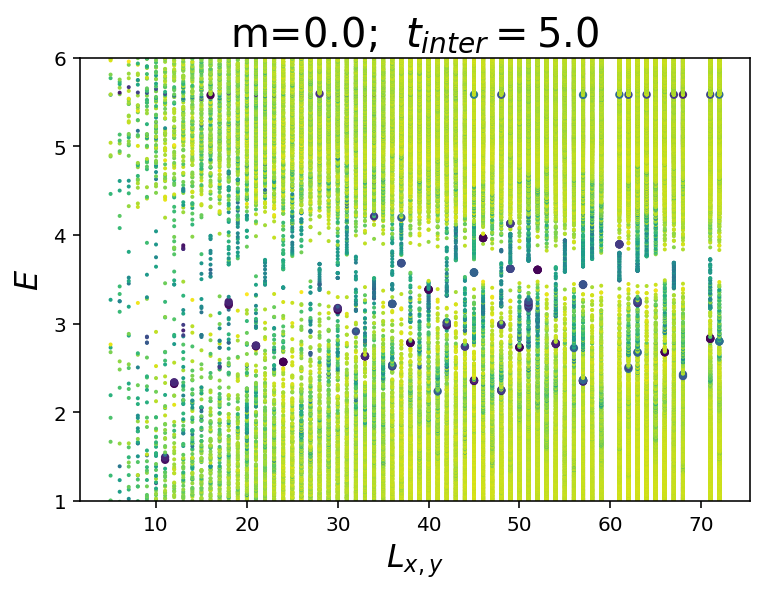}
		\subcaption{}
		\label{Loop_over_Lx_m_0.0_t_inter_5.0}
	\end{subfigure}
	\begin{subfigure}[t]{0.2\textwidth}
		\includegraphics[width=4cm,angle=0] {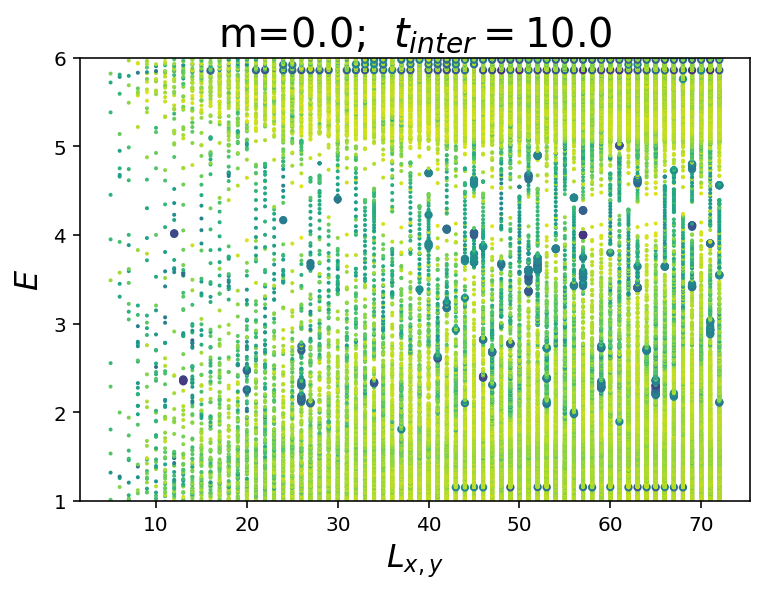}
		\subcaption{}
		\label{Loop_over_Lx_m_0.0_t_inter_10.0}
	\end{subfigure}
	\caption{Energies (vertical axis) and fractal dimensions (color) of the states plotted against the system size $L_x = L_y$ for 
$t_{inter}=0.0$ (\ref{Loop_over_Lx_m_0.0_t_inter_0.0}), $t_{inter}=1.0$ (\ref{Loop_over_Lx_m_0.0_t_inter_1.0}),
$t_{inter}=5.0$ (\ref{Loop_over_Lx_m_0.0_t_inter_5.0}) and $t_{inter}=10.0$ (\ref{Loop_over_Lx_m_0.0_t_inter_10.0}). The color represents each state's fractal dimension $D_q$ at $q=2$, and we use the same colorbar as on the Fig. \ref{Loop_over_m_and_t_inter}. 
States with fractal dimensions smaller than $0.75$ are emphasized by larger dot size. One can see that at $t_{inter}=0.0$ and $t_{inter}=1.0$, topological edge states, as well as localized states within the bulk band exist for any $L_x$.
 In the strongly coupled phase ($t_{inter}=5, 10$), one can see a new energy gap, but with increasing $L_{x, y}$ it becomes less pronounced. The most localized states (blue) may or may not lie within the energy gap. }
	\label{Loop_over_Lx}
\end{figure}

\begin{figure}
	\begin{subfigure}[t]{0.15\textwidth}
		\includegraphics[width=3cm,angle=0]{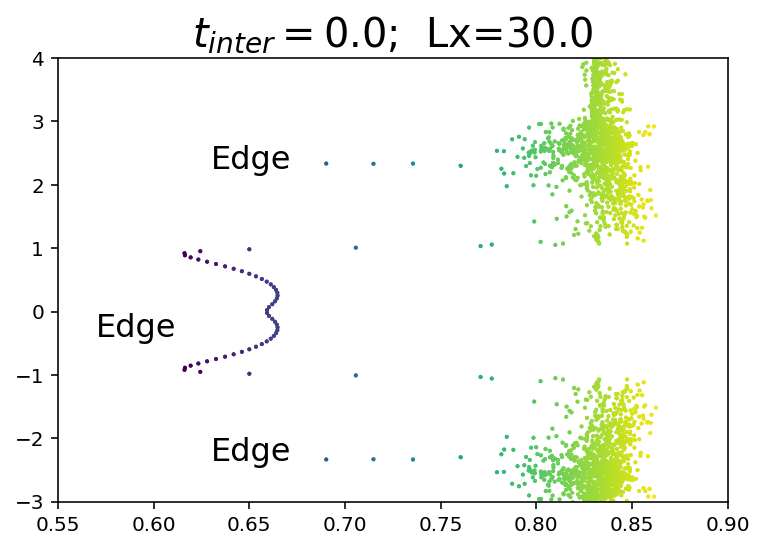}
		\subcaption{}
		\label{m_0.0_t_inter_0.0_Lx_30}
	\end{subfigure}
	\begin{subfigure}[t]{0.15\textwidth}
		\includegraphics[width=3cm,angle=0]{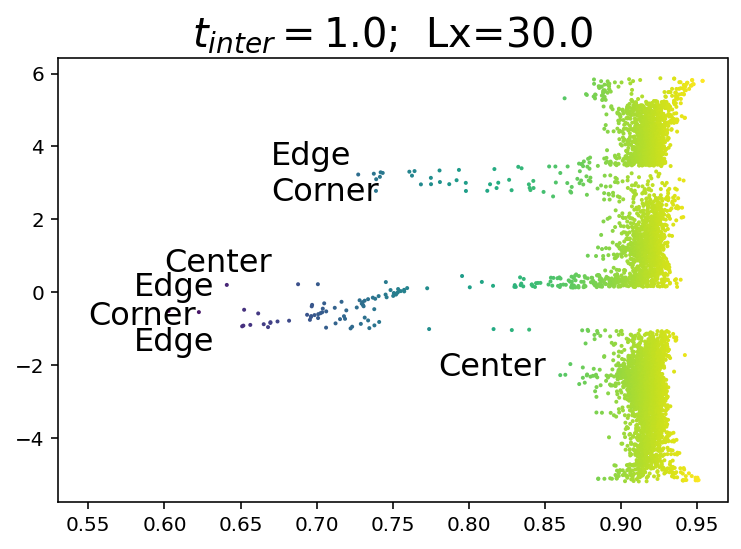}
		\subcaption{}
		\label{m_0.0_t_inter_1.0_Lx_30}
	\end{subfigure}
	\begin{subfigure}[t]{0.15\textwidth}
		\includegraphics[width=3cm,angle=0]{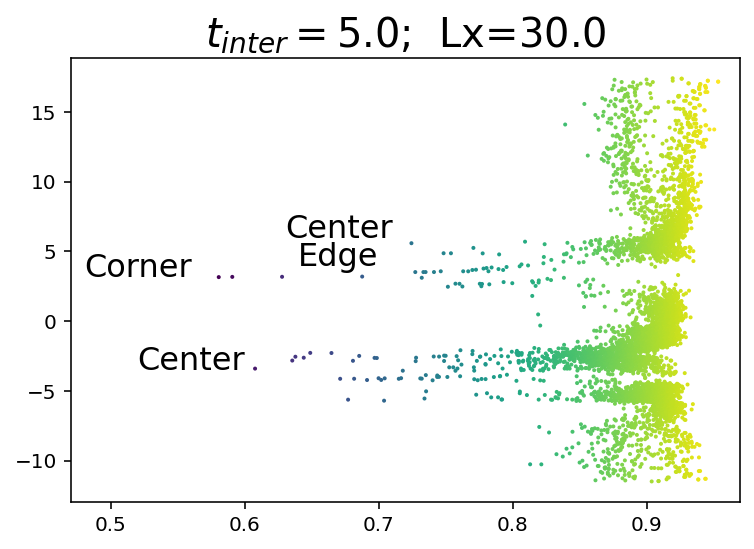}
		\subcaption{}
		\label{m_0.0_t_inter_5.0_Lx_30}
	\end{subfigure}
	\begin{subfigure}[t]{0.15\textwidth}
		\includegraphics[width=3cm,angle=0]{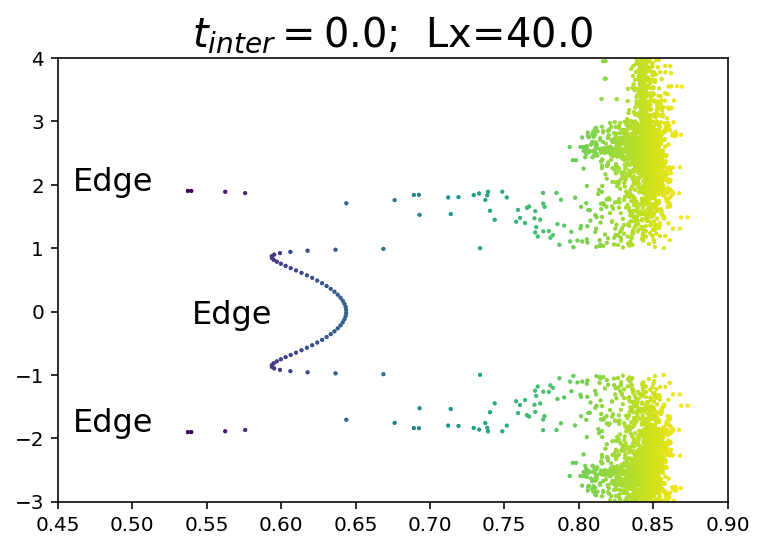}
		\subcaption{}
		\label{m_0.0_t_inter_0.0_Lx_40}
	\end{subfigure}
	\begin{subfigure}[t]{0.15\textwidth}
		\includegraphics[width=3cm,angle=0]{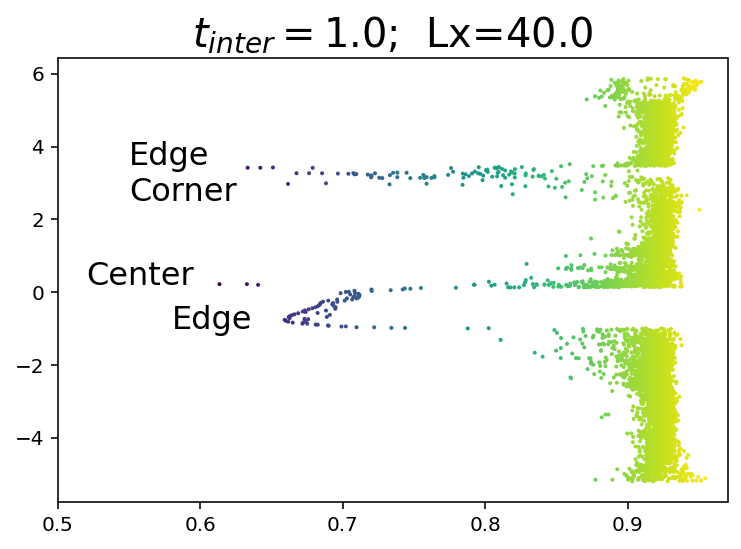}
		\subcaption{}
		\label{m_0.0_t_inter_1.0_Lx_40}
	\end{subfigure}
	\begin{subfigure}[t]{0.15\textwidth}
		\includegraphics[width=3cm,angle=0]{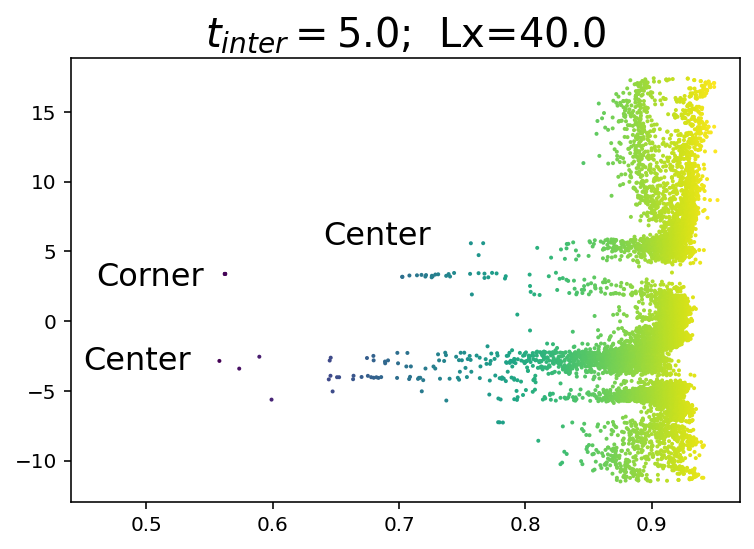}
		\subcaption{}
		\label{m_0.0_t_inter_5.0_Lx_40}
	\end{subfigure}

\caption{Energies of states plotted against their fractal dimensions for $m=0$, $L_{x, y}=30$, $t_{inter}=0$ (\ref{m_0.0_t_inter_0.0_Lx_30}), $t_{inter}=1$ (\ref{m_0.0_t_inter_1.0_Lx_30}), $t_{inter}=5$ (\ref{m_0.0_t_inter_5.0_Lx_30})
 and $L_{x, y}=40$, $t_{inter}=0$ (\ref{m_0.0_t_inter_0.0_Lx_40}), $t_{inter}=1$ (\ref{m_0.0_t_inter_1.0_Lx_40}), $t_{inter}=5$ (\ref{m_0.0_t_inter_5.0_Lx_40}). One can see that at $t_{inter}=0$ (\ref{m_0.0_t_inter_0.0_Lx_30}, \ref{m_0.0_t_inter_0.0_Lx_40}), the system forms a Chern insulator with topological edge states in the bulk gap, but in addition possesses edge states within the bulk band. As $t_{inter}$ increases (\ref{m_0.0_t_inter_1.0_Lx_30}, \ref{m_0.0_t_inter_1.0_Lx_40}), some of the edge states become localized near the corners. Such corner states persist at strong interlayer coupling (\ref{m_0.0_t_inter_5.0_Lx_30}, \ref{m_0.0_t_inter_5.0_Lx_40}). In addition, states localized at the center of the lattice emerge.}
\end{figure}

Let us look more closely at the properties of the states in our model. For simplicity, we focus on the case $m=0$. As we mentioned, at $t_{inter} = 0$, our model is a superposition of two Haldane models, which in turn host bulk states separated by an energy gap and edge states within the bulk gap.  However, we also found that our system hosts additional edge states within the bulk gap. We observe such states in the cases of both zigzag-armchair and bearded-armchair edges, but we find them to lie at slightly different energies. Specifically, in the case of zigzag-armchair edges, these edge states lie at $E\approx\pm 2.3$ (see Fig. \ref{m_0.0_t_inter_0.0_Lx_30}), whereas in the case of bearded-armchair edge, these states lie at energy $E \approx \pm 1.9$. One can see these states for various system sizes $L_{x, y}$ on the Fig. \ref{Loop_over_Lx_m_0.0_t_inter_0.0} - the bearded-armchair configuration corresponds to lattice sizes $7, 10, 13, 16 \ldots$, and zigzag-armchair is realized for all other integer values of $L_{x, y}$. 

As we increase $t_{inter}$, the edge states in the bottom `band' disappear. However, in the top `band', they persist, and a few of them tend to become localized within the corners of our system (see Figs.  \ref{m_0.0_t_inter_1.0_Lx_30}, \ref{m_0.0_t_inter_1.0_Lx_40}). Specifically for $L_{x, y}=30$, we also find that a few of the topological edge states become localized at the corners (see Fig. \ref{m_0.0_t_inter_1.0_Lx_30}; note that we could not establish connection between these modes and a shape of the lattice boundary). As we increase $t_{inter}$ further, the bulk gap closes, and the system becomes gapless. At even larger $t_{inter}$, another energy gap opens. Interestingly, we found that even though such a gap exists at any system size, its precise location and width varies a lot between different lattice sizes (see Figs.  \ref{Loop_over_Lx_m_0.0_t_inter_5.0},  \ref{Loop_over_Lx_m_0.0_t_inter_10.0}). We attribute it to the fact that since our system is quasiperiodic, the system becomes unique for each lattice size, and therefore it is extremely difficult to achieve `continuum limit' (see also discussion in the Sec. \ref{Sec:Multifractal_properties}). We however, can see that as the system size increases, this gap becomes less pronounced (see Fig. \ref{Loop_over_Lx}).

As predicted in the previous works \cite{PhysRevResearch.2.033071}, our system hosts states localized at the corners. However, we argue that their location is \textit{not} attributed to the bulk gap. Indeed, we can see that for $L_{x, y}=30$ (see Fig. \ref{Loop_over_t_inter_m_0.0_Lx_30}), the localized corner states are in the bulk gap for a range of $t_{inter}$ between $\approx 5$ and $10$, but at smaller or larger $t_{inter}$, they merge with the bulk spectrum. On the other hand, at $L_{x, y}=20$ (see Fig. \ref{Loop_over_t_inter_m_0.0_Lx_20}), we find that the corner modes lie entirely within the bulk spectrum away from the gap, but at the same time, within the bulk gap, there are extended states. We note that both $L_{x, y}=20, 30$ in our example correspond to zigzag-armchair edges of the lattice, and even have a similar structure of the corners (see Fig. \ref{Lattice_Lx_20}, \ref{Lattice_Lx_30} ). We observe qualitatively similar picture for $L_{x, y}=40$ with `armchair-bearded' edge structure (Fig. \ref{Lattice_Lx_40}): localized corner modes lie within the bulk gap at $t_{inter} \gtrsim 5$, but within the bulk states at smaller $t_{inter}$. ((see Fig. \ref{Loop_over_t_inter_m_0.0_Lx_40})).

\begin{figure}
	\hspace{-0.5cm}  
	\begin{subfigure}[t]{0.2\textwidth}
		\includegraphics[width=4cm,angle=0] 
		{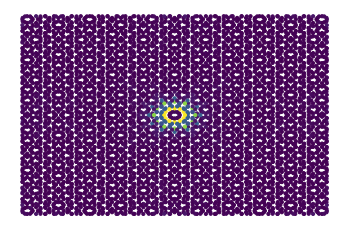}
		\subcaption{$i=1556$, $E=-3.4$}
		\label{m_0.0_t_1556}
	\end{subfigure}
	\begin{subfigure}[t]{0.2\textwidth}
		\includegraphics[width=4cm,angle=0] 
		{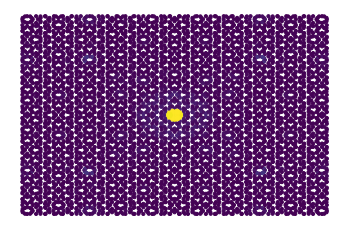}
		\subcaption{$i=2236$, $E=-2.28$}
		\label{m_0.0_t_2236}
	\end{subfigure}
	\begin{subfigure}[t]{0.2\textwidth}
		\includegraphics[width=4cm,angle=0] 
		{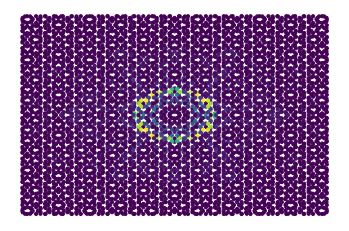}
		\subcaption{$i=1978$, $E=-2.65$}
		\label{m_0.0_t_1978}
	\end{subfigure}
	\begin{subfigure}[t]{0.2\textwidth}
		\includegraphics[width=4cm,angle=0] 
		{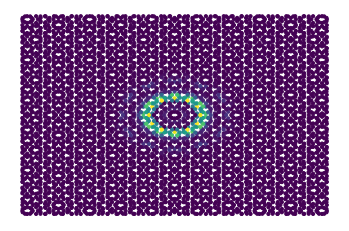}
		\subcaption{$i=2075$, $E=-2.55$}
		\label{m_0.0_t_2075}
	\end{subfigure}
	\begin{subfigure}[t]{0.2\textwidth}
		\includegraphics[width=4cm,angle=0] 
		{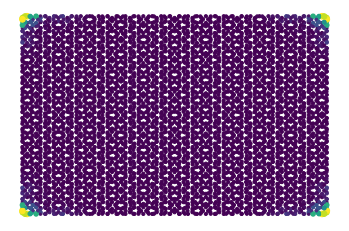}
		\subcaption{$i=4039$, $E=3.15$}	
		\label{m_0.0_t_4039}
	\end{subfigure}
	\begin{subfigure}[t]{0.2\textwidth}
		\includegraphics[width=4cm,angle=0] 
		{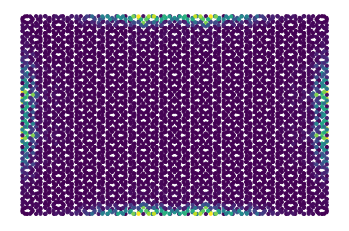}
		\subcaption{$i=4045$, $E=3.52$}	
		\label{m_0.0_t_4045}
	\end{subfigure}
\caption{$|\psi_i|^2$ for a few of the localized states at $m=0$, $t_{inter}=5$, $L_{x, y}=30$. We observed states localized at the center of the lattice (\ref{m_0.0_t_1556}, \ref{m_0.0_t_2236}), forming a ring (\ref{m_0.0_t_1978}, \ref{m_0.0_t_2075}), a well as localized at the corners (\ref{m_0.0_t_4039}) and at the edges (\ref{m_0.0_t_4045})}
\label{Localized_states}
\end{figure}

Even more interestingly, we observe numerically that our system hosts other kinds of localized states. For instance, even at large $t_{inter}$, our system still hosts states localized near edges (see Fig. \ref{m_0.0_t_4045} ). But most surprisingly, we find that, at $m=0$, the states with the smallest fractal dimensions are localized in the center of the lattice (see Fig. \ref{m_0.0_t_1556}, \ref{m_0.0_t_2236}). 

\begin{figure}
	\hspace{-0.5cm}  
	\begin{subfigure}[t]{0.2\textwidth}
		\includegraphics[width=4cm,angle=0] 
		{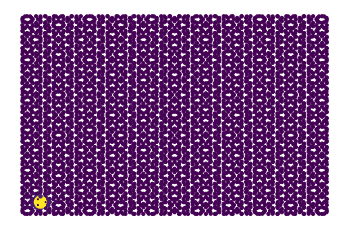}
		\subcaption{$i=2760$, $E=1.3$}
		\label{m_1.5_t_2760}
	\end{subfigure}
	\begin{subfigure}[t]{0.2\textwidth}
		\includegraphics[width=4cm,angle=0] 
		{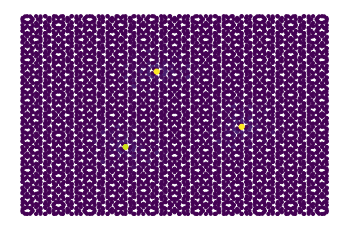}
		\subcaption{$i=2762$, $E=1.4$}
		\label{m_1.5_t_2762}
	\end{subfigure}
	\begin{subfigure}[t]{0.2\textwidth}
		\includegraphics[width=4cm,angle=0] 
		{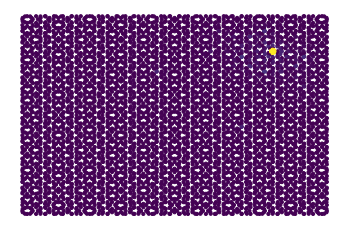}
		\subcaption{$i=2764$, $E=1.4$}
		\label{m_1.5_t_2764}
	\end{subfigure}
	\begin{subfigure}[t]{0.2\textwidth}
		\includegraphics[width=4cm,angle=0] 
		{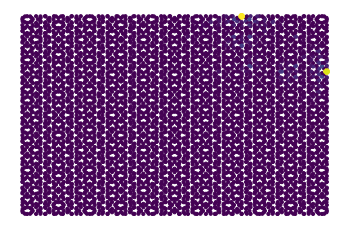}
		\subcaption{$i=2843$, $E=1.8$}
		\label{m_1.5_t_2843}
	\end{subfigure}
	\begin{subfigure}[t]{0.2\textwidth}
		\includegraphics[width=4cm,angle=0] 
		{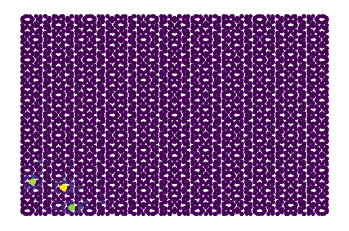}
		\subcaption{$i=2845$, $E=1.85$}	
		\label{m_1.5_t_2845}
	\end{subfigure}
	\begin{subfigure}[t]{0.2\textwidth}
		\includegraphics[width=4cm,angle=0] 
		{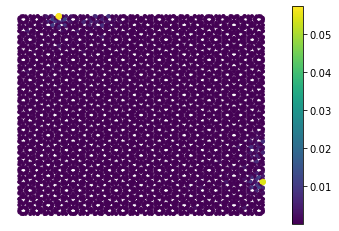}
		\subcaption{$i=2854$, $E=1.9$}	
		\label{m_1.5_t_2854}
	\end{subfigure}
\caption{$|\psi_i|^2$ for a few of the most localized localized states at $m=1.5$, $t_{inter}=5$, $L_{x, y}=30$. One can see that the states can be localized at various locations of the lattice. }
\label{Localized_states_m_1.5}
\end{figure}

We also find that at $m \ne 0$, both corner and `center' modes disappear, but instead our system hosts states localized at other specific locations in the bulk of the sample (see Fig. \ref{Localized_states_m_1.5}). We attribute their existence to the lack of translational symmetry. Indeed, in periodic crystals, isolated localized states cannot exist within the bulk of the material, but this does not have to be the case in a material which lacks translational symmetry. The same thing from a different perspective: in a periodic crystal, corner modes can be viewed as a consequence of bulk polarization \cite{PhysRevB.96.245115}, which is uniform in the bulk. However, bulk polarization does not have to be uniform in a system without translational symmetry. We leave the problem of building a more comprehensive theory of localized states in quasicrystals for future work.

\subsection{Multifractal properties}

\label{Sec:Multifractal_properties}

\begin{figure}
	\includegraphics[width=8cm,angle=0] 
	{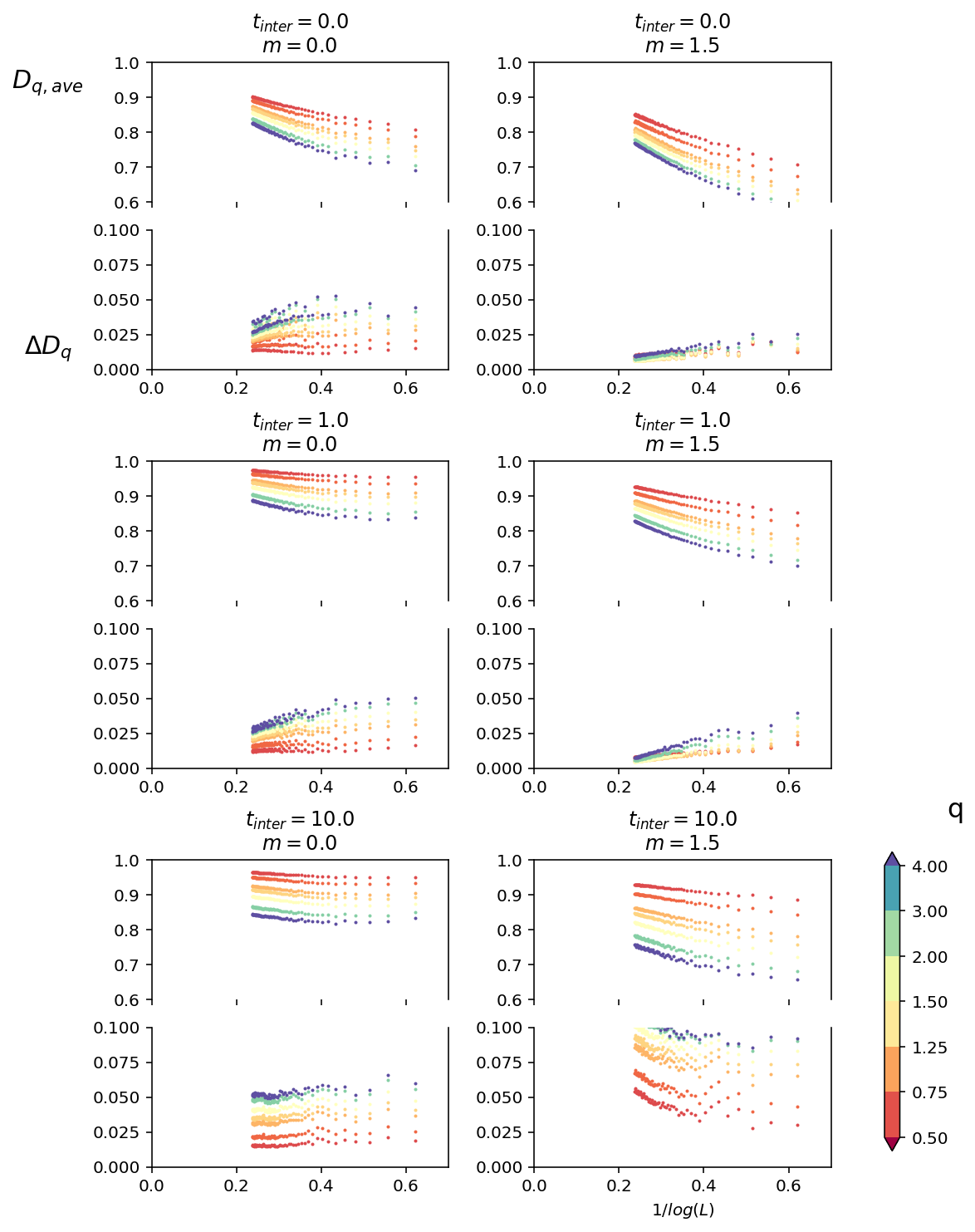}
\caption{ Average fractal dimensions $D_{q, ave}$ and their standard deviations $\Delta D_q$ over all states in a system plotted against its size $L_x$ (we assume $L_x=L_y$). We choose a list of $q = 0.5, 0.75, 1.25, 1.5, 2.0, 3.0, 4.0$. In the weakly coupled phase, $D_{q, ave}$ converges to $1$, and $\Delta D_q$ converges to $0$ - this means that all states are extended. In the strongly coupled phase, both $D_{q, ave}$ and $\Delta D_q$ converge to values between $0$ and $1$ - the system is multifractal. }
\label{Plot_energies_and_average_fractal_dim_loop_over_L}
\end{figure}

To explore multifractal properties of our model, we plot average fractal dimension and its standard deviation for all states against the system size (once again we assume that the system forms a square with $L_x = L_y$) and a few values of $q$ (Fig. \ref{Plot_energies_and_average_fractal_dim_loop_over_L}). At zero $t_{inter}$, we observe that as the system size increases, the average fractal dimension of all states in the system approaches $1$ and its standard deviation approaches $0$. In other words, we confirm that in both topological and non-topological weakly coupled phases, the bulk eigenstates are fully extended as they are expected to be in a fully periodic system. 

\begin{figure}
	\includegraphics[width=8cm,angle=0] 
	{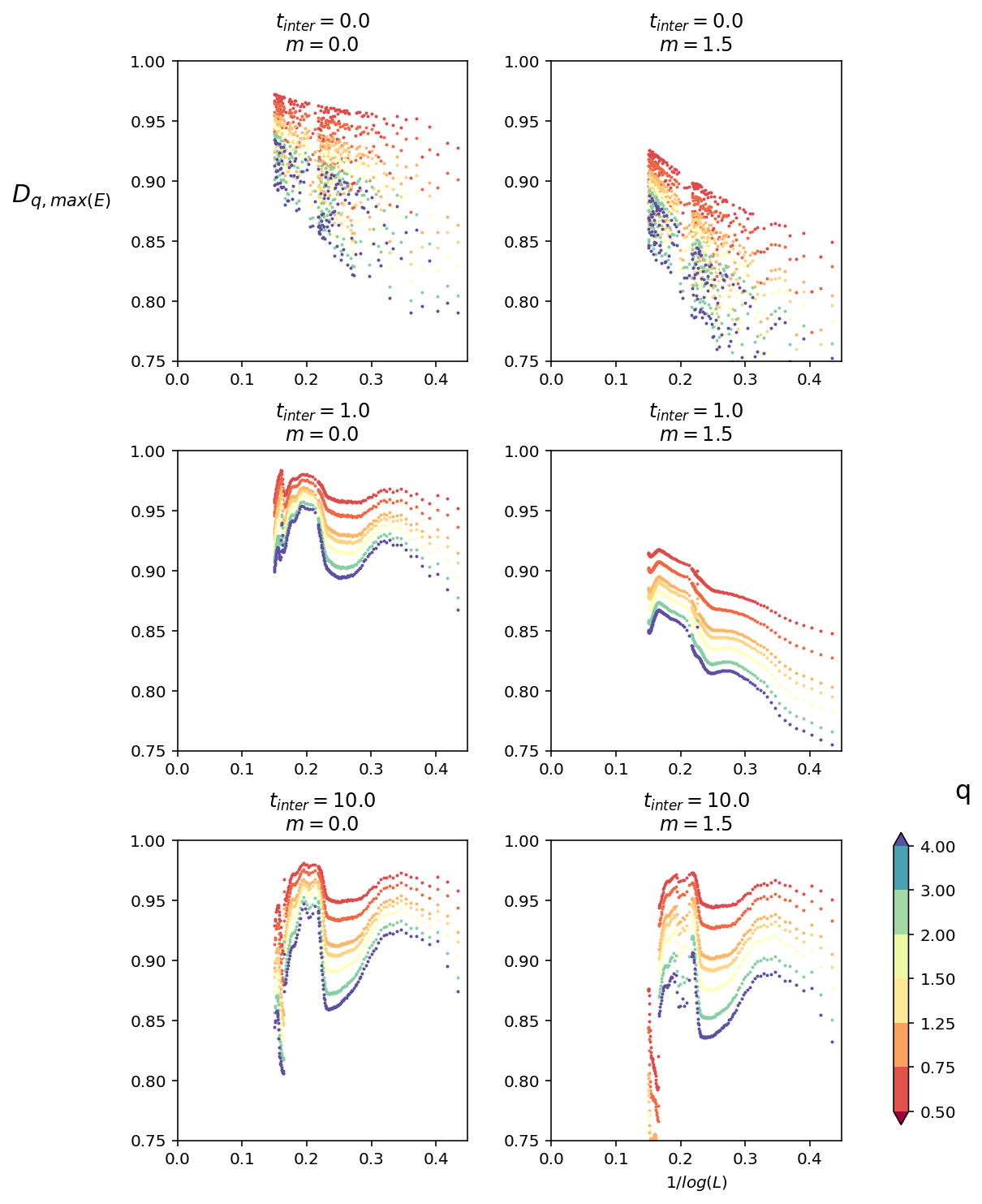}
\caption{Fractal dimensions $D_q$ of the states with the largest energies plotted against $1/\log(L_x)$ for $q = 0.5, 0.75, 1.25, 1.5, 2.0, 3.0, 4.0$. We assume $L_x=L_y$. The wavefunctions were obtained using Lanczos diagonaliztion. We see that for $t_{inter}=0$, $D_q$ approaches $1$ with increasing $L_x$, but does not approach $1$ for $t_{inter} \ne 0$.}
\label{Lanczos_loop_over_L}
\end{figure}

Characterizing multifractal properties at non-zero $t_{inter}$ turns out to be challenging because finite size corrections to fractal dimensions scale as $1/\log(L)$ for a system with a linear size $L$ \footnote{One can see this just by writing a wavefunctions as $\psi \sim C/\sqrt{L}$ and plugging into the Eq. \ref{Frac_dim}}, and because of that it is impossible to obtain reliable results for any realistic system size, which can be handled numerically. For this reason, we cannot say confidently whether average fractal dimensions approach $1$ at $t_{inter}=1$. However, when we plot average fractal dimensions and their deviations as a function of the system size in the 'strongly coupled' phase ($t_{inter}=10$), we observe that the former converge to values smaller than $1$, and the latter converge to values greater than $0$. Furthermore, we can see that as the system size grows, average fractal dimensions converge to different values for various $q$. Thus we can say fairly confidently that eigenstates in the strongly coupled phase are multifractal - they are neither extended, nor localized.

\begin{figure}
	\begin{subfigure}[t]{0.15\textwidth}
		\includegraphics[width=3cm,angle=0]{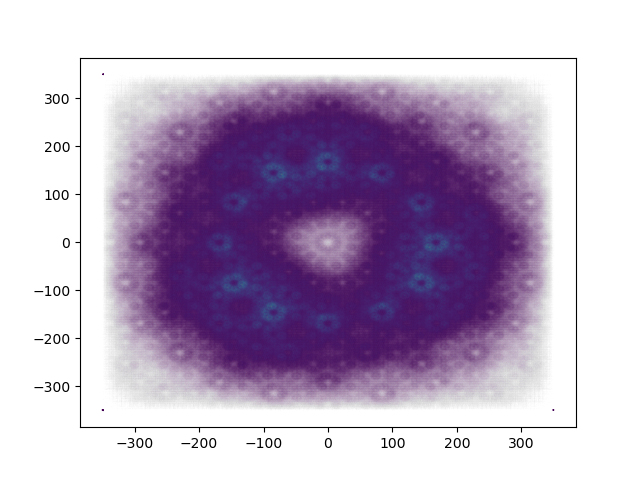}
		\subcaption{}
		\label{m_1.5_t_inter_10.0_Lx_350}
	\end{subfigure}
	\begin{subfigure}[t]{0.15\textwidth}
		\includegraphics[width=3cm,angle=0]{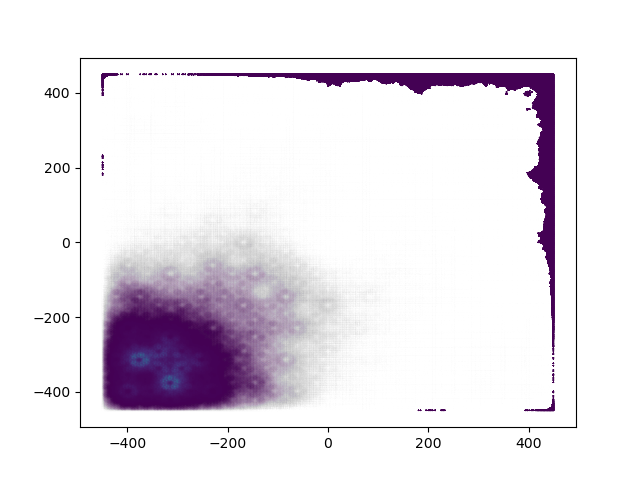}
		\subcaption{}
		\label{m_1.5_t_inter_10.0_Lx_450}	
	\end{subfigure}
\caption{ $|\psi_i|^2$ of the states with highest energies obtained by Lanczos diagonalization for $m=1.5$, $t_{inter}=10$, $L_{x, y}=350$ (\ref{m_1.5_t_inter_10.0_Lx_350}) and $L_{x, y}=450$ (\ref{m_1.5_t_inter_10.0_Lx_450}).  }
\label{Lanczos_wavefunctions}
\end{figure}

To complement our understanding of multifractaility in our system, we also study it using Lanczos diagonalization. This method makes it possible to consider much larger system sizes (up to $L_{x, y} \approx 800$). Specifically, for each $L_{x,y}$, we obtain eigenstates with the largest energies and compute their fractal dimensions (see Fig.  \ref{Lanczos_loop_over_L}). As previously, we can see that for $t_{inter}=0$, i.e. for periodic system, fractal dimension of the `top' state approaches $1$ with increasing system size. On the other hand, for $t_{inter} \ne 0$, we can see that the fractal dimension of the `top' state does not approach $1$. However, despite reaching huge lattice sizes, we are still not able to ensure that they are sufficient to make predictions about infinite-size properties. For instance, at $m=1.5$, $t_{inter}=10$, we observe that the wavefunction of the `top' state remains rotationally-symmetric for sizes up to $L_{x, y} \approx 420$, whereas at larger sizes, starts getting localized near one of the corners (see Fig. \ref{Lanczos_wavefunctions}). We believe that this is yet another manifestation of the fact that in a quasicrystalline system, continuum limit is reached extremely slowly, and even system with the largest sizes that can be handled by modern computers may still be far from it. In the future, it would be of interest to find means to study multifractality in our model analytically.

\subsection{Topological properties}

\label{Sec:Topological properties}

\subsubsection{Topological entanglement entropy}

\label{Sec:Topological_entanglement_entropy}

To study entanglement properties of a many body system, one has to partition it into two subsystems in coordinate space. For a subsystem $A$ with density matrix $\rho_A$, the entanglement entropy is defined as $S = - \mathrm{tr} \rho_A \log \rho_A$. If a 2D system has smooth boundary, its entanglement entropy scales linearly with system size $L$ as
\begin{eqnarray}
S = \alpha L - \gamma,
\end{eqnarray}
where the constant $\gamma$ is a known as a topological entanglement entropy - a term present only in topological systems. We note that in this definition, it is important that the system's boundary is smooth, otherwise its entanglement entropy might get additional contributions e.g. due to corners (\cite{PhysRevB.94.125142, PhysRevB.99.155153}).

\begin{figure}
		\includegraphics[width=6cm,angle=0] 
		{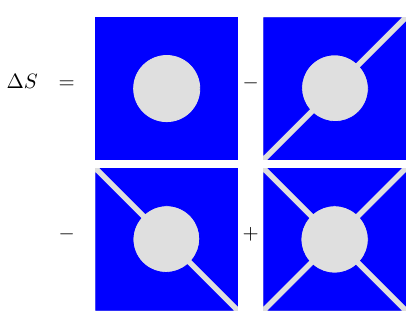}
\caption{A scheme of partitioning a system, which we use to compute the topological contribution $\gamma$ to the entanglement entropy. Each of the circular curves has a radius of $L_x/2$. We compute entanglement entropy for each of the figures and then use this substraction scheme to isolate the contribution from the topological term $\gamma$. }
\label{Entanglement_figures_cropped}
\end{figure}

To extract the topological contribution $\gamma$, we compute the difference between entanglement entropies of several subsystems in such a way that their leading term, which is linear over $L$, as well as subleading contributions due to corners exactly cancel out. To do this, we choose a configuration of four systems shown on the Fig. \ref{Entanglement_figures_cropped}. We compute entanglement entropies of each subsystem using the method suggested in Refs. \cite{Helmes:2016ahb, Ingo_Peschel_2003, PhysRevB.99.155153}, which we briefly describe in Sec. \ref{Sec:Calculation_of_the_entanglement_entropy}.

We plot the results for $\gamma$ as functions of $m$, $t_{inter}$ on the Figs. \ref{Gamma_loop_over_t_inter} (we assume $E_F = 0$). We can see that in the 'weakly coupled topological' phase, $\gamma$ is a non-zero constant, which in turn confirms that it is a topological quasicrystal. In a similar way, in the 'weakly coupled non-topological' phase, $\gamma$ is equal to zero, which means that the system is non-topological. Once the gap closes, the topological entanglement entropy becomes ill-defined. 

\begin{figure}
\includegraphics[width=8cm,angle=0]{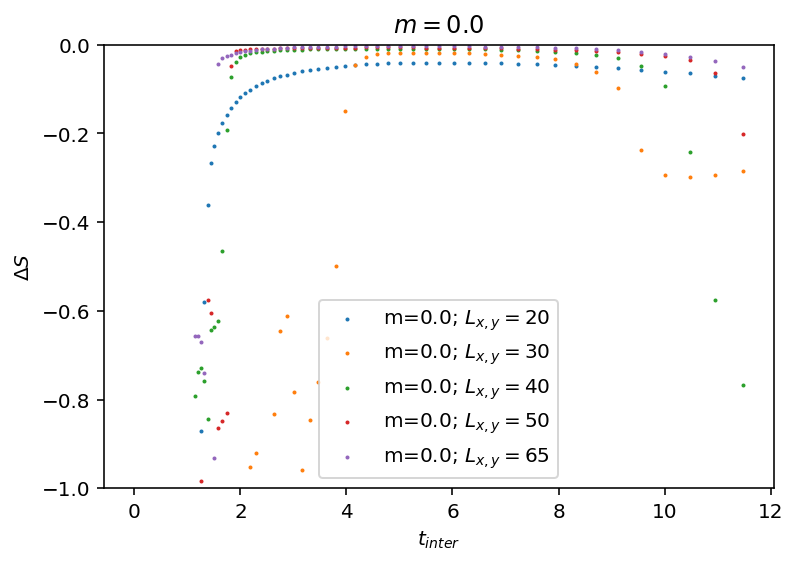}
\caption{Topological entanglement entropy $\gamma$ plotted against $t_{inter}$ for $m=0$, 
in the case when the Fermi level lies within the bulk energy gap of the `strongly coupled' system. We can see that with increasing system size $L_x$ (we assume $L_x = L_y$), it approaches zero.
\label{Gamma_large_t_inter}
}
\end{figure}

We also compute $\gamma$ for large $t_{inter}$ at $m=0$, assuming that the Fermi level $E_F$ lies within the energy gap, which we previously found on the Figs \ref{Loop_over_t_inter_m_0.0_Lx_20}, \ref{Loop_over_t_inter_m_0.0_Lx_30}, \ref{Loop_over_t_inter_m_0.0_Lx_40}. We obtain that at large system size, $\gamma$ becomes close to zero (see Fig. \ref{Gamma_large_t_inter}). This suggests that our `strongly coupled' gapped system is not topological.

\begin{figure}
	\hspace{-3cm}
	\begin{subfigure}[t]{0.2\textwidth}
		\includegraphics[width=8cm,angle=0] 
		{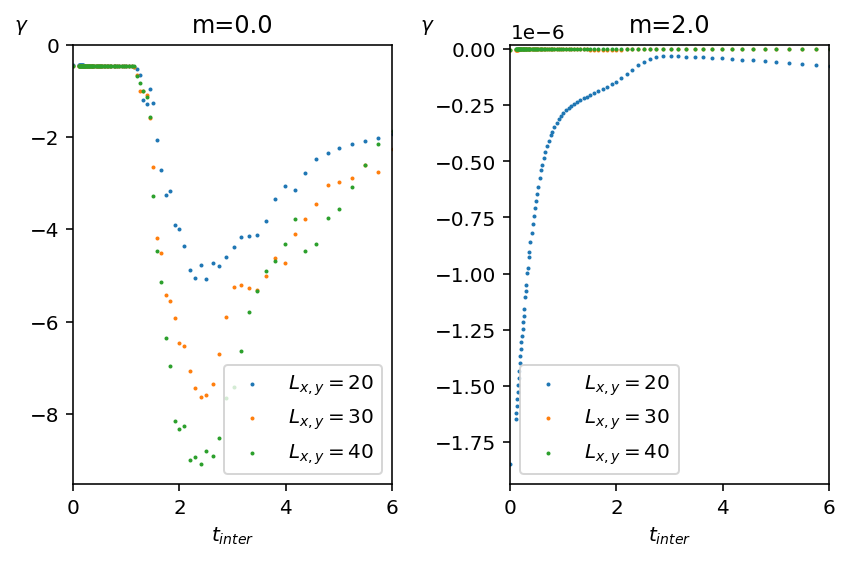}
		\subcaption{}	
		\label{Gamma_loop_over_t_inter}
	\end{subfigure}
	\\
	\hspace{-3cm}
	\begin{subfigure}[t]{0.2\textwidth}
		\includegraphics[width=8cm,angle=0] 
		{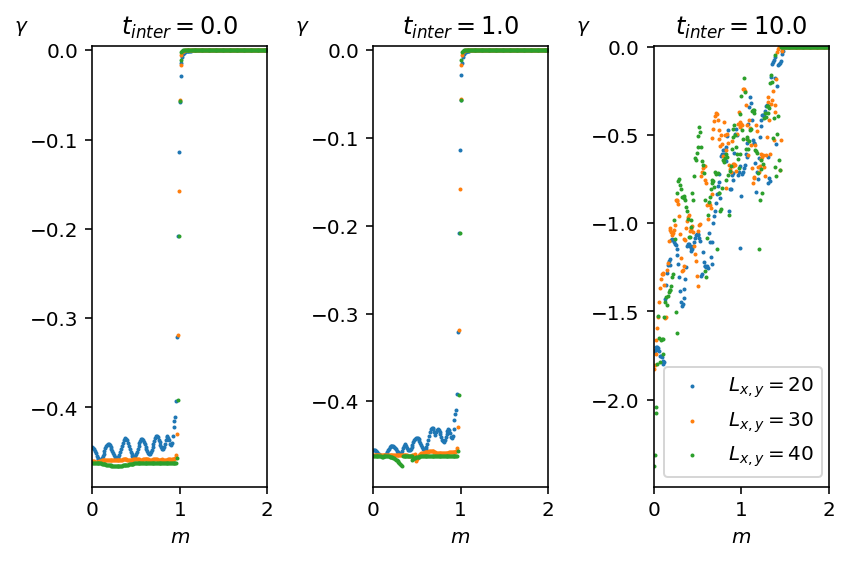}
		\subcaption{}	
		\label{Gamma_loop_over_m}
	\end{subfigure}
\caption{Topological entanglement entropy $\gamma$ plotted against $t_{inter}$ (\ref{Gamma_loop_over_t_inter}) for $m=0, 2$ and against $m$ (\ref{Gamma_loop_over_m}) for $t_{inter}= 0.0, 1.0, 10.0$. Here we assume $E_F=0$. It is equal to a non-zero constant in the topological gapped phase and zero in the non-topological phase}
\label{Gamma_loop_over_t_inter}
\end{figure}

\subsubsection{Local Chern marker}

\label{Sec:Local_Chern_marker}

It was argued in Ref \cite{PhysRevB.84.241106}, that Chern number is not well-defined for a system of finite size because 
if one defines it in the same way as the corresponding infinite-size expression, the answer becomes exactly zero.  
 Instead it was suggested that topological properites of a system can be characterized empirically using 'local Chern marker' defined in the following way. First, one can define a projector to the subspace of filled 
 states as $ P(\vec{r}_i, \vec{r}_j) = \sum\limits_{E_{f} < E_F} \psi_{f}(\vec{r}_i) \psi^{\dagger}_{f}(\vec{r}_j) $ (here the indices $i, j$ numerate the lattice sites, and the summation is performed over filled states numerated by an index $f$).
Next, one may project 
the coordinate operators as 
$ \tilde{X}(\vec{r}_i , \vec{r}_j) = \sum\limits_{\vec{r}_k} P(\vec{r}_i, \vec{r}_k) x_k P(\vec{r}_k, \vec{r}_j) $
and $ \tilde{Y}(\vec{r}_j , \vec{r}_i) = \sum\limits_{\vec{r}_{k'}} P(\vec{r}_j, \vec{r}_{k'}) y_{k'} P(\vec{r}_{k'}, \vec{r}_i) $. The local Chern marker is defined as
\begin{eqnarray}
C(\vec{r}_i) = \frac{2 \pi i}{S}  \sum_{\vec{r}_j} \left[ \tilde{X} (\vec{r}_i, \vec{r}_j) , \tilde{Y} (\vec{r}_j, \vec{r}_i)  \right],
\label{Local_Chern_marker}
\end{eqnarray}
where $S$ is an average area per each lattice site on the lattice. One may check that for conventional Haldane model on a finite lattice, this local Chern marker inside the bulk is equal to the value of Chern invariant in the corresponding continuum model, whereas near the edges, its value is very large and has opposite magnitude in such a way that its sum over all lattice sites is zero: $\sum\limits_i C(\vec{r}_i) = 0$.

\begin{figure}
	\hspace{-3cm}
	\begin{subfigure}[t]{0.2\textwidth}
		\includegraphics[width=8cm,angle=0] 
		{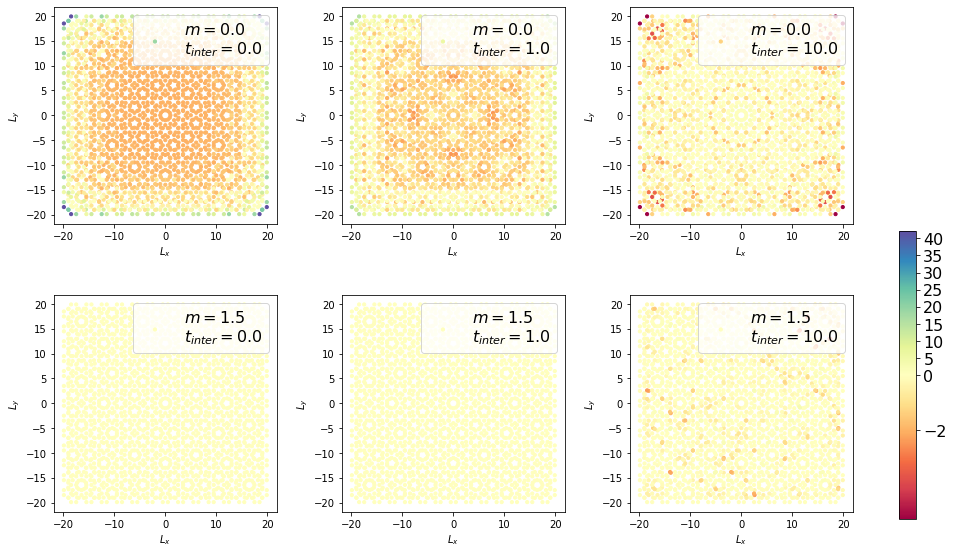}
		\subcaption{}	
		\label{Local_Chern_marker_lattice}
	\end{subfigure}
	\\
	\hspace{-3cm}
	\begin{subfigure}[t]{0.2\textwidth}
		\includegraphics[width=6cm,angle=0] 
		{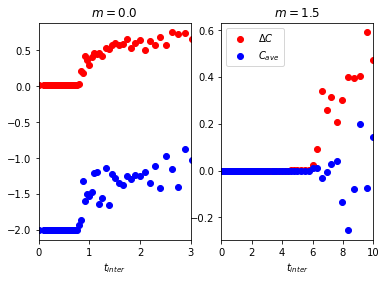}
		\subcaption{}	
		\label{Local_Chern_marker_loop_over_t_inter}
	\end{subfigure}
\caption{ \ref{Local_Chern_marker_lattice} Local Chern marker plotted for lattices with $-L_{x,y} < x, y < L_{x,y}$ at various values of $m$, $t_{inter}$ and $L_x = L_y=30$. One can see that in the bulk of a topological phase, it is approximately equal to Chern invariant, and in the non-topological phase, it is equal to zero. In the multifractal phase, it fluctuates within the lattice. \ref{Local_Chern_marker_loop_over_t_inter} Averaged local Chern marker $C_{ave}$ and its standard deviation $\Delta C$ taken over the inner part of the lattice with $-L_{x, y}/2 < x,y < L_{x, y}/2$. It remains constant and approximately equal to the bulk Chern invariant in the weakly coupled phase, but starts fluctuating at large $t_{inter}$.} 
\label{Local_Chern_markers}
\end{figure}

We plot the local Chern marker of the $30^{\circ}$ twisted bilayer Haldane model at $E_F=0$ for various values of $t_{inter}$ on Fig. \ref{Local_Chern_marker_lattice}. One may see that in the 'weakly coupled' phases, it behaves in the same way as in the case of two non-coupled copies of Haldane model on a monolayer. Specifically, in the 'weakly coupled topological phase', its bulk value remains the same and equal to the value in the case of zero interlayer coupling. Similarly, in the 'weakly coupled non-topological' phase, its bulk value is equal to zero. On the other hand, at larger values of $t_{inter}$, the local Chern marker starts behaving differently from the case of periodic lattice. 
Specifically, the local Chern marker no longer has a form of constant value in the bulk and large, but opposite value at the edge. Instead, regions with $C(\vec{r}_i)$ values of opposite signs start emerging within the bulk of the system, and sufficiently far away from the transition point, these regions become indistinguishable from the edges of the system. In other words, at large $t_{inter}$, the local Chern marker $C(\vec{r}_i)$ strongly fluctuates within the bulk. so that there is no distinction between the bulk and the edges as in a crystalline topological insulator.

\begin{figure}
\includegraphics[width=8cm,angle=0]{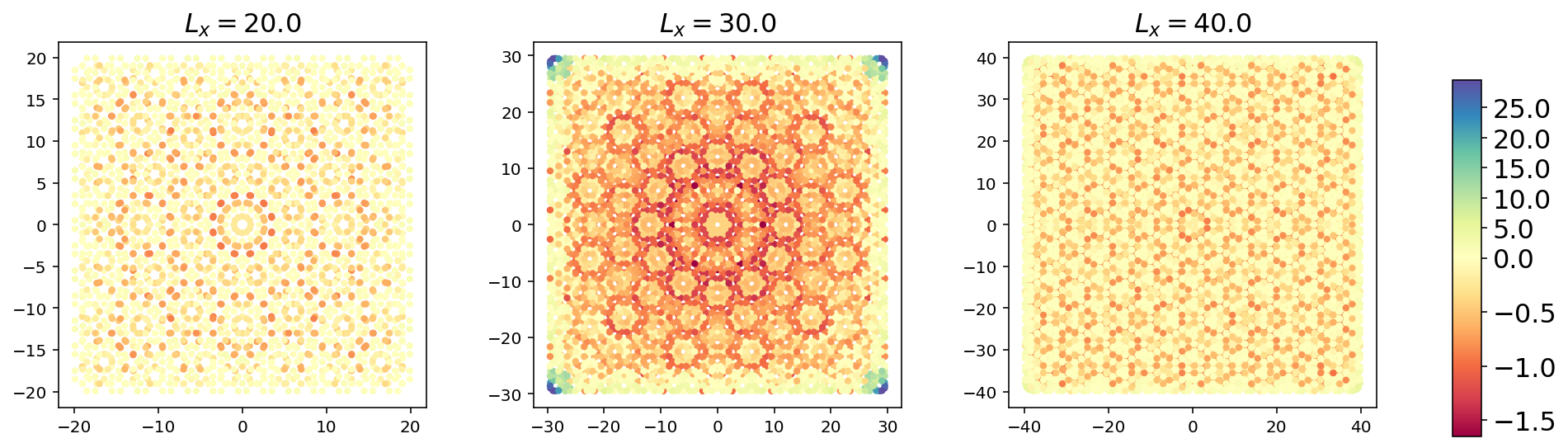}
\caption{Local Chern marker $C(\vec{r}_i)$ plotted for the system in the strongly coupled phase ($t_{inter}=10$) in the case when the Fermi level lies within the bulk energy gap. We assume $m=0.0$ and consider lattice sizes $L_{x, y} = 20, 30, 40$. One can see that $C(\vec{r}_i)$ fluctuates within the bulk, which in turn means that the system does not have well-defined topological properties.}
\label{Local_Chern_markers_large_t_inter}
\end{figure}

For completeness, we also plot the local Chern marker in the strongly coupled phase ($t_{inter}=10$) in the case when $E_F$ lies within the bulk gap (see Fig. \ref{Local_Chern_markers_large_t_inter}). We can see once again that it strongly fluctuates within the bulk. Thus, we conclude that the strongly coupled gapped phase cannot be characterized by a well-defined behavior of the local Chern marker, which is in turn just another evidence that this phase is non-topological. 



\subsubsection{Anomalous Hall conductivity}
\label{Sec:Anomalous_Hall_conductivity}

We start our analysis from introducing electric current of our model. Specifically, if a tight-binding Hamiltonian for a given system has hoppings $H_{ij}$ between sites $i, j$, then in the presence of electric field they change according to Peierls substitution and thus become $H_{ij} e^{\frac{ie}{\hbar}\int\limits_{\vec{r}_i}^{\vec{r}_j} \vec{A} \vec{dr}}$, where $\vec{A}$ is a vector potential at the bond connecting the sites at positions $\vec{r}_i$ and $\vec{r}_j$. If we expand these hoppings to the first order in $\vec{A}$, we obtain that the electric current operator has the following expression 
\begin{eqnarray}
 \vec{J}_{ij} = \frac{ie}{\hbar}  H_{ij} (\vec{r}_j - \vec{r}_i).
\label{Electric_current}
\end{eqnarray}
Here $\vec{J}_{ij}$ is a vector directed along the bond between the sites $i, j$. In fact, the last expression could be immediately obtained by writing Heisenberg equation and assuming that the current is proportional to the velocity operator. 

Further, one can compute a response to applied electric field by using an expression for the electric current (\ref{Electric_current})  and applying Kubo formula. Anomalous Hall conductivity would be its antisymmetric part, and thus it would 
be given by the Eq. (\ref{AHE_Kubo_formula}).
After lengthy, but straightforward calculations (specifically writing electrons Green functions and performing Matsubaba summation; see Sec. \ref{Sec:KuboFormulaAHE} for more details), the expression for anomalous Hall conductivity can be brought to the following form
\begin{eqnarray}
\sigma^{(xy)}_{ij}
&=& - \frac{e^2}{2 \hbar^2 } \sum\limits_{f - filled \atop e - empty} \sum\limits_{k, l} \frac{1}{(E_f - E_e)^2}
\label{AHE_gen_expr}
\\
&&\times
\left\{
\psi^{\dagger}_{e, i} H_{ij} (x_j - x_i) \psi_{f, j}
\cdot  \psi^{\dagger}_{f, k} H_{kl} (y_l - y_k) \psi_{e, l}
\right. 
\nonumber\\
&&
\left.
 - \psi^{\dagger}_{f, i} H_{ij} (x_j - x_i) \psi_{e, j}
\cdot  \psi^{\dagger}_{e, k} H_{kl} (y_l - y_k) \psi_{f, l}
\right. 
\nonumber\\
&& -
\left.
\psi^{\dagger}_{e, i} H_{ij} (y_j - y_i) \psi_{f, j}
\cdot  \psi^{\dagger}_{f, k} H_{kl} (x_l - x_k) \psi_{e, l}
\right. 
\nonumber\\
&&
\left.
 + \psi^{\dagger}_{f, i} H_{ij} (y_j - y_i) \psi_{e, j}
\cdot  \psi^{\dagger}_{e, k} H_{kl} (x_l - x_k) \psi_{f, l}
\right\}.
\nonumber
\end{eqnarray}
In this equation, the indices $i, j$ numerate the lattice sites between which the electric current is computed, and $k, l$ numerate the sites over which summation is performed. In addition, the summation is performed over filled states numerated by an index $f$ and empty states numerated by an index $e$. Specifically, $E_{f}, E_e$ are energies of the filled and empty states respectively, and similarly $\psi_{f, i} \equiv \psi_f(\vec{r}_i)$ and $\psi_{e, i} \equiv \psi_e(\vec{r}_i)$ are the wavefunctions of the filled and empty states respectively at the lattice site $i$.

Let us emphasize once again, that in the Eq. (\ref{AHE_gen_expr}), $\sigma_{ij}$ describes electric current through a bond between the lattice sites $i, j$ (we assumed that the electric field is uniform). However, one can sum electric currents through all bonds next to a given lattice site $i$ and consider $\sigma_i  = \sum\limits_j \sigma_{ij}$. For a crystalline system, one can take a step further and split the lattice site index $i$ into two indices $i', \alpha$ numerating unit cells and sublattice degrees of freedom respectively. The well-known integer value of the anomalous Hall conductivity is obtained after an additional summation over $\alpha$ (see Sec. \ref{Sec:AHE_crystalline}). A simple explanation of the latter is as follows: for an infinite crystalline system one may transform any dependency over unit cell index $i'$ into momentum representation, and the well-known integer AHE occurs at zero momentum. However, its expression in terms of Berry curvature still contains summation over sublattice degrees of freedom. Nevertheless, on a honeycomb lattice, the two sublattice degrees of freedom are equivalent to each other, and therefore the AHE for each of them would be the same, i.e. half-integer. 

More interestingly, properties of AHE obtained from the Eq. (\ref{AHE_gen_expr}) on a finite lattice 
are fundamentally different from the case of an infinite crystal. Namely AHE on a finite lattice behaves in a similar way to a local Chern marker: its sum over all lattice sites $i, j$ is equal to zero, its bulk value is constant for a crystalline lattice and equal to the value it would have in a corresponding infinite lattice, and its value at the edges has opposite sign and large magnitude so that its sum over all sites near the eges exactly cancels out the sum over all sites in the bulk.  At first sight, this may seem paradoxal, but there is a simple argument for it. Indeed it is well-known that on an infinite lattice, AHE is proportional to an integer topological invariant, and phases with different topological invariants cannot be smoothly transformed into each other, but are separated by phase transitions. However, it is also well-known that phase transitions do not exist on a finite lattice, but appear only in thermodynamic limit. This means that topologically distinct phases cannot exist on a finite lattice! In other words, any phase on a finite lattice is topologically equivalent to trivial. For this reason, if one sums the local Chern marker (\ref{Local_Chern_marker}) over all lattice sites $r_i$, or one sums AHE (\ref{AHE_gen_expr}) over all sites $i, j$, one obtains exactly zero.

\begin{figure}
	\hspace{-3cm}
	\begin{subfigure}[t]{0.2\textwidth}
		\includegraphics[width=8cm,angle=0] 
		{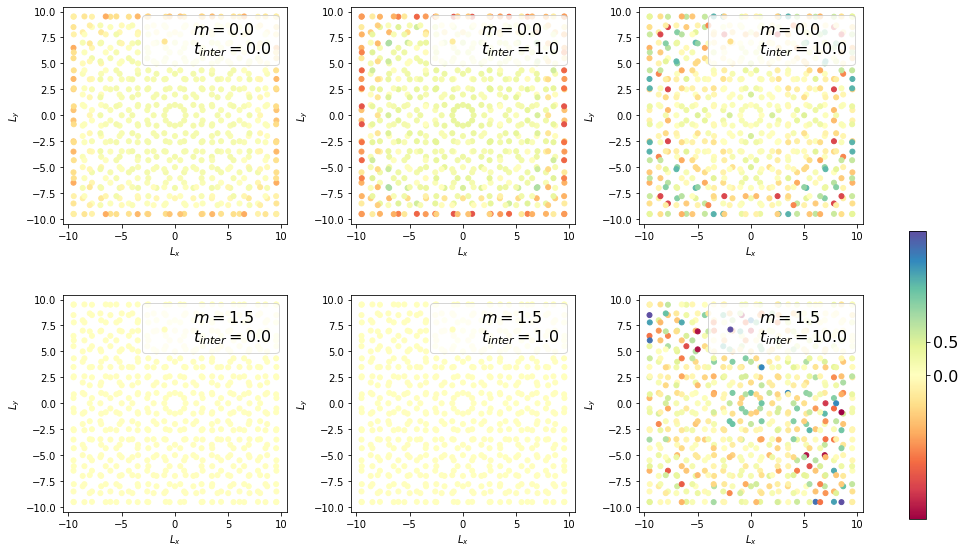}
		\subcaption{}	
		\label{Sum_Hall_conductivities_plot}
	\end{subfigure}
	\\
	\hspace{-3cm}
	\begin{subfigure}[t]{0.2\textwidth}
		\includegraphics[width=6cm,angle=0] 
		{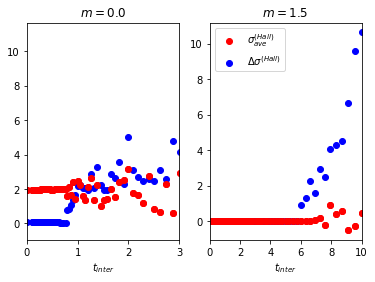}
		\subcaption{}	
		\label{Sum_Hall_conuctivities_loop_over_t_inter}
	\end{subfigure}
\caption{ \ref{Sum_Hall_conductivities_plot} Hall conductivity $\sigma_i$ (in units of $e^2/\hbar^2$) at each site $i$ plotted for lattices with $-L_{x, y} < x, y < L_{x, y}$ at various values of $m$, $t_{inter}$ and $L_x = L_y = 30$. One can see that in the bulk of a topological phase, it is approximately equal to $0.5$ - Chern invariant divided by the number of sites in a unit cell in each layer, and in the non-topological phase, it is equal to zero. In the multifractal phase, it fluctuates within the lattice. \ref{Sum_Hall_conuctivities_loop_over_t_inter} Averaged $\sigma_i$ summed over the four sublattice degrees of freedom ($A/B$ and top/bottom) and its standard deviation taken over the inner part of the lattice with $-L_{x, y}/2 < x,y < L_{x, y}/2$. It remains constant and approximately equal to the bulk Chern invariant in the weakly coupled phase, but starts fluctuating at large $t_{inter}$.} 
\label{Sum_Hall_conductivities}
\end{figure}

We present our numerical calculations of  $\sigma_i$ for our model of twisted bilayer on the Fig. \ref{Sum_Hall_conductivities}. One can see that in the case of zero interlayer coupling, i.e. when the model is just a superposition of two periodic layers, $\sigma_i$ behaves just like a local Chern marker. In the 'weakly coupled topological phase', it has a constant value inside the bulk equal to the value it would have in the limit of infinite size, whereas at the boundary its value has an opposite sign and large magnitude so that the total sum $\sum\limits_i \sigma_i$ is zero. Once the interlayer coupling $t_{inter}$ is turned on, the behavior remains the same while $t_{inter}$ is small. However, once $t_{inter}$ is increased, $\sigma_i$ is no longer uniform inside the bulk. The behavior of $\sigma_i$ shows a phase transition at the same value of $t_{inter}$ as the local Chern marker, and as $t_{inter}$ increases further, the distinction between the bulk and the edge behavior smoothly disappears. This fact supports our argument that for a quasicrystal there is no distinction between bulk and edge as in a crystal, but instead there are 'bulk-like' and 'edge-like' regions.

Finally let us explain the meaning of zero total $\sum\limits_i \sigma_i$ from the experimental perspective. Indeed, introducing the vector potential $\vec{A}$ into our Hamiltonian with open boundary conditions is equivalent to placing our sample under external electric field. However, if one places a finite and isolated sample of a material with non-trivial AHE under electric field, there would be no electric currents flowing into or out of the sample because it does not have any contacts, through which the current might flow. In other words, our calculations 
physically mean that if a finite and isolated sample is placed under electric field, there will appear electric currents inside the sample, but no currents into or out of it. In order to obtain a physically measurable anomalous Hall conductivity, one has either to consider explicitly the flow of electric current through contacts, or just to assume that the sample is infinite in the direction of electric current, as it was done in multiple past works  (e.g. \cite{10.1007/3-540-46637-1_2}). Since within our model, the case of an infinite sample is fundamentally different from a finite one, our results can be applied only to the latter. Nevertheless, we anticipate that if our model is realized experimentally, then at zero as well as small interlayer coupling, while the system is gapped, the total Hall conuctivity would be just an integer, i.e. the sum of two Hall conductivities of the isolated samples. On the other hand, in the gapless phase, the Hall conductivity will become non-universal, and possibly it may not even have a well-defined limit at infinite system size. The latter fact can be interpreted in a way that 'strong quasicrystallinity' is qualitatively similar to strong disorder: in its presence, all observable properties become sample-dependent.

\section{Discussion}

\label{Sec:Discussion}

In this work, we have studied topological properties of quasicrystalline 30 degrees twisted bilayer honeycomb lattice. We started from the case of two uncoupled layers each of which forms Haldane model, and tracked its evolution with the increasing interlayer coupling. We found that at small interlayer coupling, 
our system exactly retains topological properties of the two uncouped layers: bulk gap and topological edge states persist, topological entanglement entropy remains constant, local Chern marker and anomalous Hall conductivity behave in exactly the same way as in the case of two uncoupled layers. This fact is not obvious because topological properties of Haldane model are characterized by Chern invariant defined in momentum space, which is in turn well-defined only in the presence of crystalline translational symmetry. In this regard, one could be concerned whether translational symmetry breaking might break topological protection, but we established that 
at small, but non-zero interlayer coupling, this is not the case: our bilayer has exactly the same topological properties as a superposition of two monolayers. In this sense, we demonstrated that two weakly coupled copies Chern insulators can be viewed as a quasicystalline topological insulator. In other words, we proposed to create a quasicystalline topological insulator by smoothly evolving crystalline topological insulator and breaking translational symmetry. 


 
We found that as the interlayer coupling increases, our system undergoes a phase transition into a gapless phase, and its topological edge states disappear. As the interlayer coupling increases even further, another bulk gap may open. We found that such a stronly coupled phase is overall multifractal, but it possesses multiple states localized at various locations of the lattice. For example, at $m=0$, our system possesses states localized at the center of the lattice, as well as its corners. However, we believe that these corner modes are not topological in nature because their energies are not tied to the bulk gap, and also because topological entanglement entropy of the strongly coupled gapped system is almost zero. More rigorous topological classification of quasicrystals is a challenging and unresolved task, which we leave for future work.

We believe that closing of the topological band gap and the disappearance of edge modes at large interlayer coupling is natural to expect. In fact, one may consider an analogy between our model and a disordered topological insulator. At weak disorder, the system has exactly the same topological properties as the topological insulator without disorder. However, once the disorder becomes strong, fluctuations of the disorder potential effectively behave like edges themselves, and this fact qualitatively explains the gap closing and implies that the bulk becomes indistinguishable from the physical edges of the system.



It would be of interest to realize our model in experiments, but it is challenging because typically (e.g. in twisted bilayer graphene), at large twist angles, interlayer coupling is very weak. However, we expect that our system should host qualitatively similar properties not only at $30^{\circ}$ twist, but at any incommensurate twist angle. Moreover, one could create quasiperiodicity not only by an incommensurate twist angle, but also by stacking together two monolayers with different lattice constants. The latter is plausible to achieve considering the fact that the family of 2D materials by now is very rich (e.g. graphene, transition metal dichalcocenides etc). One may also try to change interlayer spacing by applying high pressure between the layers. Finally, we would like to point out the possibility to realize our model using ultracold atoms (see e.g. \cite{Jotzu2014}), which have been proved to be a powerful tool to realize various physical models.

In summary, we proposed a way to realize topological quasicrystals by starting from topological crystalline materials and breaking their translational symmetries. Moreover we proposed a model of a quasicrystal, which explicitly hosts a topological phase transition. We demonstrated that non-trivial topological properties of quasicrystals can be characterized not only by local Chern markers, but also by topological entanglement entropy and anomalous Hall conductivity. Most surprisingly, we established that the anomalous Hall conductivity in an open system, in our case in a quasicrystal behaves qualitatively in the same way as a local Chern marker. We hope that in the future, our results may be used to obtain rigorous topological classification of quasicrystals. We are also interested in studying their unusual transport properties and possibilities for future applications.




\section{Acknowledgements}

The author would like to thank Institute of Basic Science (Daejeon, S. Korea), in which this project started and Profs. Moon Jip Park, Kyong Min Kim, Sergej Flach, Sergey Syzranov, Predrag Nikolic, Renat Sabirianov, Luis Santos, Ivan Khaymovich, Niccolo Traverso Ziani, Matteo Carrega for helpful discussions. 
Financial support by the National Science Foundation through EPSCoR RII Track-1: Emergent Quantum Materials and Technologies (EQUATE), Award OIA-2044049 is acknowledged.

\appendix

\section{The limit of $30^{\circ}$ twisted bilayer graphene}
\label{Sec:twisted_bilayer_graphene}

\begin{figure}
	\begin{subfigure}[t]{0.15\textwidth}
		\includegraphics[width=3cm,angle=0]{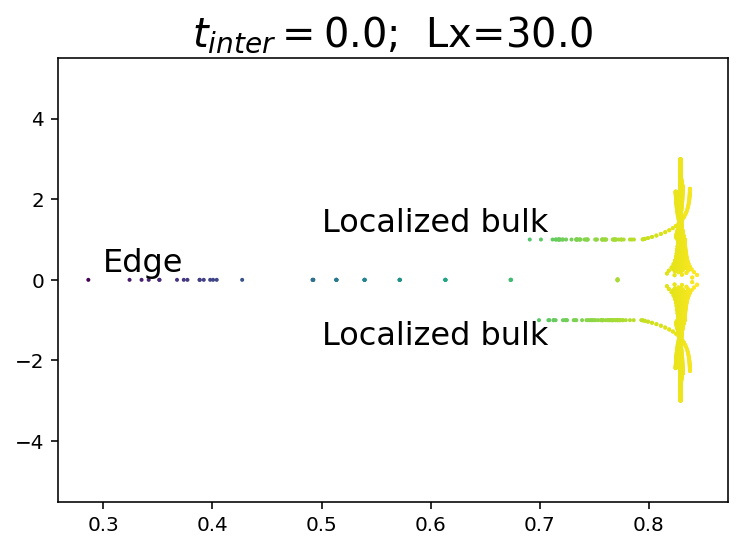}
		\subcaption{}
		\label{Graphene_m_0.0_t_inter_0.0_Lx_30}
	\end{subfigure}
	\begin{subfigure}[t]{0.15\textwidth}
		\includegraphics[width=3cm,angle=0]{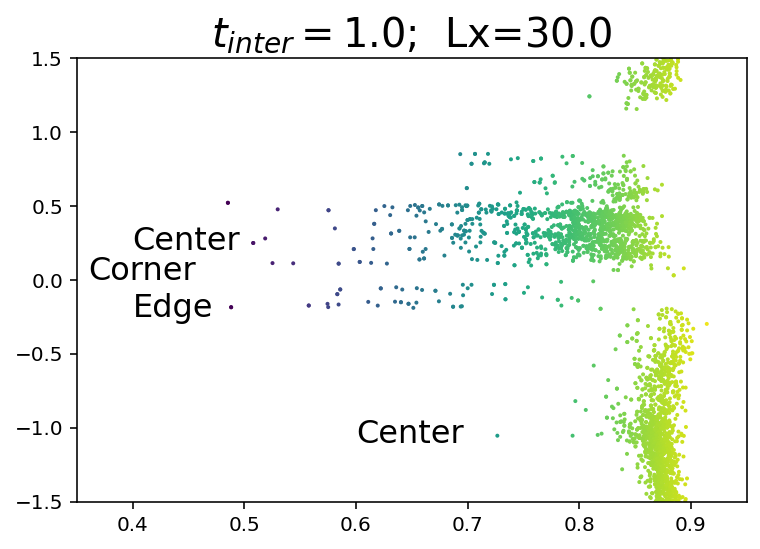}
		\subcaption{}
		\label{Graphene_m_0.0_t_inter_1.0_Lx_30}
	\end{subfigure}
	\begin{subfigure}[t]{0.15\textwidth}
		\includegraphics[width=3cm,angle=0]{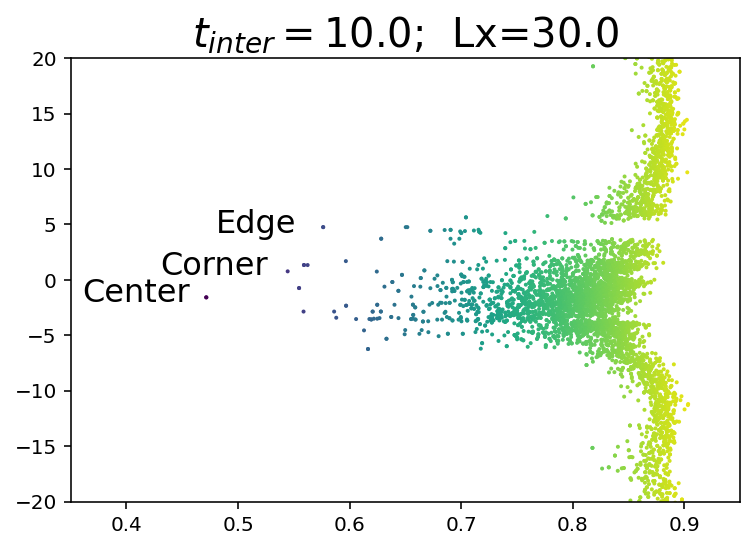}
		\subcaption{}
		\label{Graphene_m_0.0_t_inter_10.0_Lx_30}
	\end{subfigure}
	\begin{subfigure}[t]{0.15\textwidth}
		\includegraphics[width=3cm,angle=0]{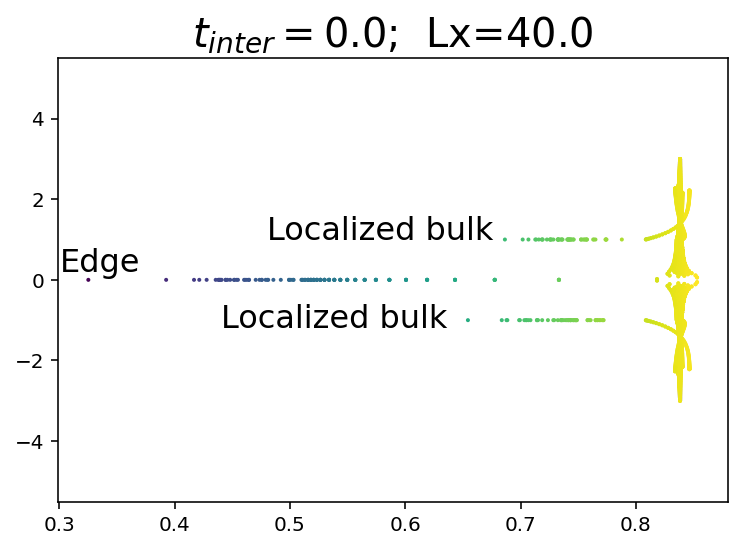}
		\subcaption{}
		\label{Graphene_m_0.0_t_inter_0.0_Lx_40}
	\end{subfigure}
	\begin{subfigure}[t]{0.15\textwidth}
		\includegraphics[width=3cm,angle=0]{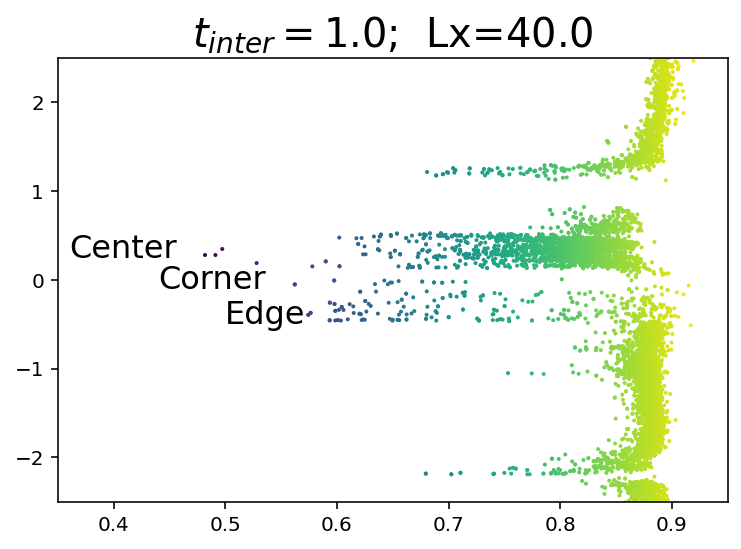}
		\subcaption{}
		\label{Graphene_m_0.0_t_inter_1.0_Lx_40}
	\end{subfigure}
	\begin{subfigure}[t]{0.15\textwidth}
		\includegraphics[width=3cm,angle=0]{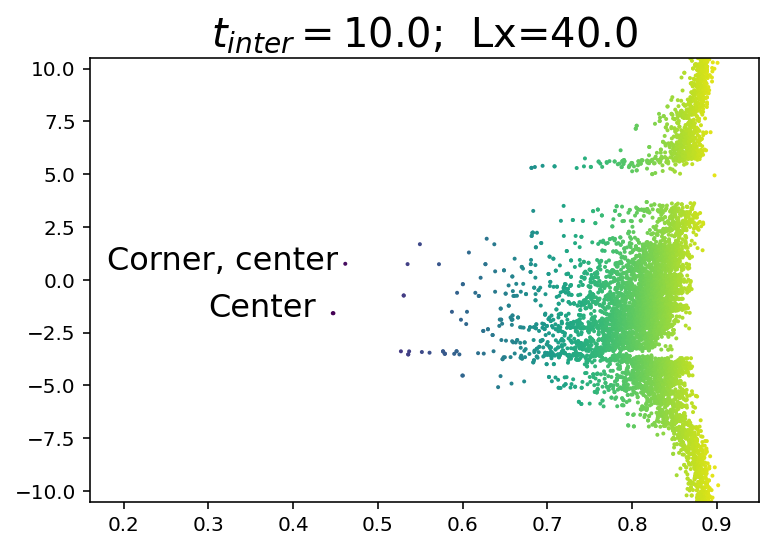}
		\subcaption{}
		\label{Graphene_m_0.0_t_inter_10.0_Lx_40}
	\end{subfigure}

\caption{Energies of states plotted against their fractal dimensions in $30^{\circ}$ twisted bilayer graphene (i.e. at $t_2 = 0$) for $m=0$, $L_{x, y}=30$, $t_{inter}=0$ (\ref{Graphene_m_0.0_t_inter_0.0_Lx_30}), $t_{inter}=1$ (\ref{Graphene_m_0.0_t_inter_1.0_Lx_30}), $t_{inter}=10$ (\ref{Graphene_m_0.0_t_inter_10.0_Lx_30})
 and $L_{x, y}=40$, $t_{inter}=0$ (\ref{Graphene_m_0.0_t_inter_0.0_Lx_40}), $t_{inter}=1$ (\ref{Graphene_m_0.0_t_inter_1.0_Lx_40}), $t_{inter}=10$ (\ref{Graphene_m_0.0_t_inter_10.0_Lx_40}). One can see that at $t_{inter}=0$, the system hosts armchair edge states at zero energy, as well as localized bulk states due to band structure degeneracy. At $t_{inter} \ne 0$, the system hosts various localized states including states localized at the corners and at the center of the lattice respectively. We observe that the location of these states is not attributed to the bulk gap.}
\end{figure}

It is known that Haldane model describes graphene in the limit when only the nearest neighbors couplings are present, i.e. at $t_2 = 0$. Graphene's energy spectrum is gapless, yet is possesses `armchair` edge states at zero energy (\cite{PhysRevB.84.195452, Wakabayashi_2010}). Numerically at $t_{inter} = 0$, we observe edge states at zero energy at any kind of lattice termination. In addition, we observe states localized along one direction within the bulk at $E = \pm 1$ (see Fig. \ref{Graphene_m_0.0_t_inter_0.0_Lx_30}, \ref{Graphene_m_0.0_t_inter_0.0_Lx_40}). The latter fact can be explained by writing graphene dispersion in momentum space as 
\begin{eqnarray}
E = \pm \sqrt{  \sin^2 \left( \frac{\sqrt{3} k_y}{2} \right) 
+ \left(  \cos \left( \frac{\sqrt{3} k_y}{2} \right)    +   2\cos\left(\frac{k_x}{2}\right)  \right)^2
 }
 \nonumber\\
\end{eqnarray}
(this equation can be obtained by writing the Hamiltonian for an isolated layer (\ref{H_alpha}) at $t_{intra}=1$,  $t_2=0$, $m=0$ and transforming it into momentum representation)
and noticing that at $E = \pm 1$, its Fermi surface is hexagonal. In other words, at $E = \pm 1$, for the same value of $k_x$, there are many states with different $k_y$, which in turn means that states can become localized in $y$ direction. 

\begin{figure}
	\includegraphics[width=7cm,angle=0]
	{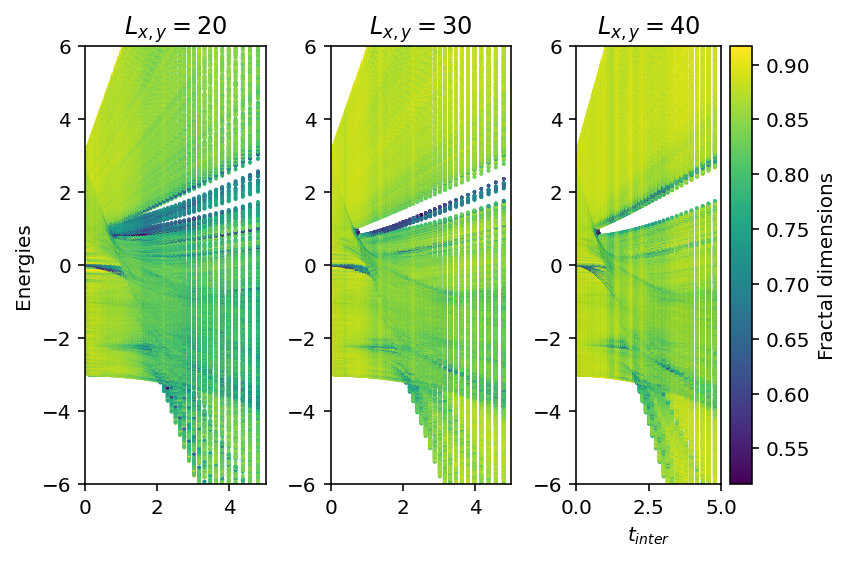}
\caption{Energies of states in $30^{\circ}$ twisted bilayer graphene (i.e. the model from the Eqs. (\ref{MainHamiltonian}-\ref{V_1_2}) with $t_{intra}=1$, $t_2=0$, $m=0$, $r_0=1$, $r_{max}=2$) plotted against $t_{inter}$. We consider lattice sizes $L_x=L_y = 20, 30, 40$. The color represents the state's fractal dimensions computed for $q=2$, as shown on the colorbar. We observe that at small non-zero $t_{inter}$, the edge states get finite energy width. At large $t_{inter}$, the edge states disappear, and a bulk energy gap opens.}
\label{Graphene_loop_over_t_inter}
\end{figure}

Now let us explore the evolution of the energy spectrum with increasing interlayer coupling. We observe that at finite $t_{inter}$, the edge states get a finite energy width and in addition, get more localized (see Fig. \ref{Graphene_loop_over_t_inter}). We also observe that a few of them turn into corner states. Finally we observe the emergence of new localized states at the center of the lattice (see Fig. \ref{Graphene_m_0.0_t_inter_1.0_Lx_30} - \ref{Graphene_m_0.0_t_inter_10.0_Lx_40}).

As the interlayer coupling increases even further, the edge states disappear, but instead a bulk gap opens  (see Fig. \ref{Graphene_loop_over_t_inter}). At $L_{x, y} =20, 30$, we observe multiple in-gap states localized near edges. However, we observe corner states, as well as states localized at the center of the lattice at energies far away from the bulk gap (see Fig. \ref{Graphene_m_0.0_t_inter_10.0_Lx_30}, \ref{Graphene_m_0.0_t_inter_10.0_Lx_40}).

\begin{figure}
	\hspace{-3cm}
	\begin{subfigure}[t]{0.2\textwidth}
		\includegraphics[width=7cm,angle=0] 
		{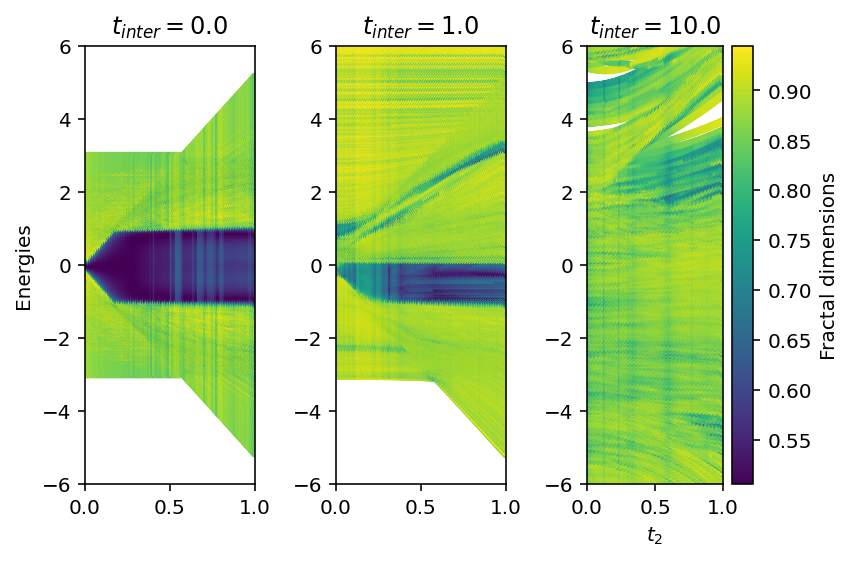}
		\subcaption{$L_{x,y}=20$}	
		\label{Loop_over_t2_Lx_20}
	\end{subfigure}
	\\
	\hspace{-3cm}
	\begin{subfigure}[t]{0.2\textwidth}
		\includegraphics[width=7cm,angle=0] 
		{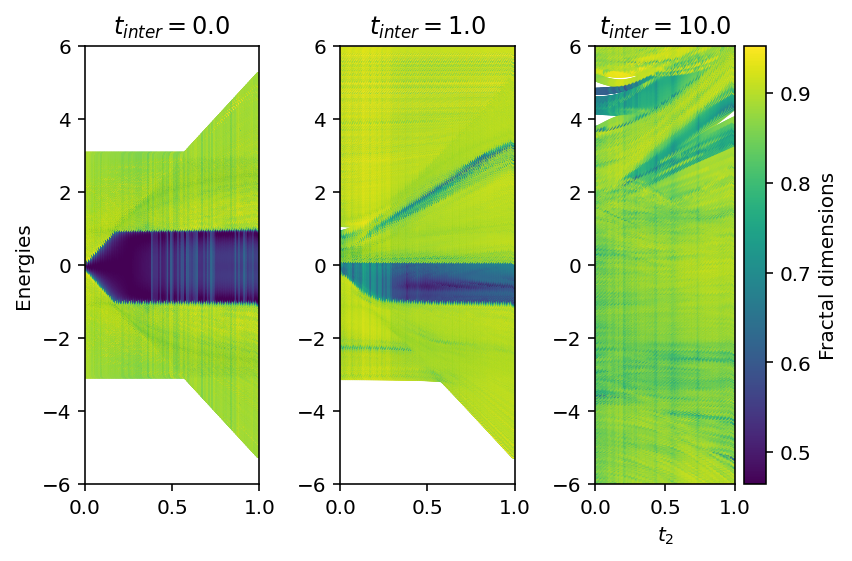}
		\subcaption{$L_{x,y}=30$}	
		\label{Loop_over_t2_Lx_30}
	\end{subfigure}
	\\
	\hspace{-3cm}
	\begin{subfigure}[t]{0.2\textwidth}
		\includegraphics[width=7cm,angle=0] 
		{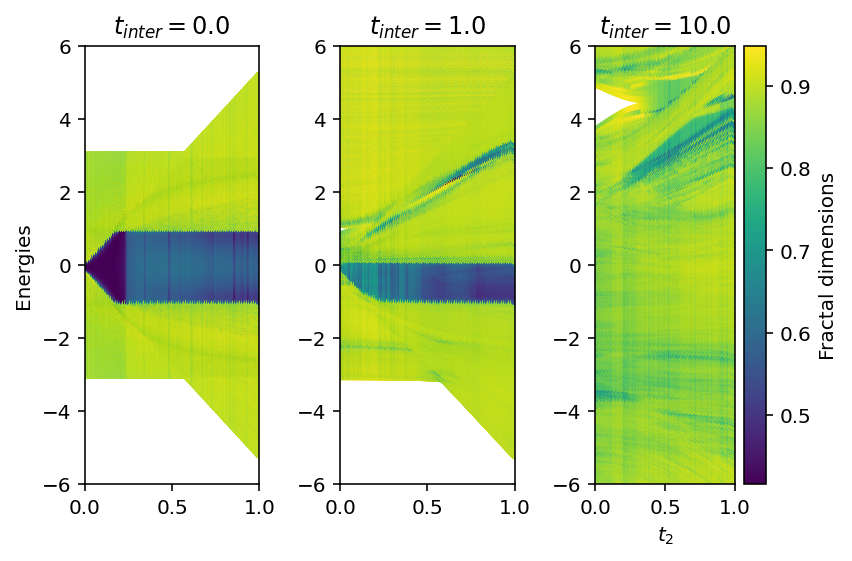}
		\subcaption{$L_{x,y}=40$}	
		\label{Loop_over_t2_Lx_40}
	\end{subfigure}
\caption{Energies of states in the model determined by Eqs. (\ref{MainHamiltonian} - \ref{V_1_2}) plotted against $t_2$. We assume $t_{intra}=1$, $m=0$, $r_0=1$, $r_{max}=2$. We also assume that the system has a shape of a square with $L_x=L_y$ and fix $L_{x,y}=20$ (\ref{Loop_over_t2_Lx_20}), $L_{x,y}=30$ (\ref{Loop_over_t2_Lx_30}) and $L_{x,y}=40$ (\ref{Loop_over_t2_Lx_40}). We observe that at $t_{inter}=0$ the bulk gap closes at $t_2=0$, at $t_{inter}=1$, the bulk gap is decreasing at $t_2 \to 0$ and at $t_{inter}=10$, the bulk gap closes and then reopens again.  
} 
\label{Loop_over_t2}
\end{figure}

For completeness, we also plot the dispersion as a function of $t_2$ for a fixed value of $t_{inter}$ (see Fig. \ref{Loop_over_t2}). Specifically, at $t_{inter} = 0$, we just observe that with decreasing $t_2$, the bulk gap gets closed. At $t_{inter} = 1$, the bulk gap decreases with decreasing $t_2$, but does not close completely. Finally, the evolution becomes more non-trivial at large $t_{inter}$. Specifically at $t_{inter}=10$, we observe that as $t_2$ decreases from $1$ to $0$, the bulk gap closes and then reopens again. In other words, we observe that the bulk gap at $30^{\circ}$ twisted bilayer graphene is not the same as the bulk gap in $30^{\circ}$ twisted Haldane model considered in the main text of the paper.     

In summary, we have demonstrated that $30^{\circ}$ twisted bilayer graphene, similarly to the $30^{\circ}$ twisted bilayer Haldane model, hosts various states localized at the corners or at the center of the lattice. However, their energies are not attributed to the bulk gap, which opens at large interlayer coupling. This is another manifestation of the fact that the corner states are non-topological. 

\begin{figure}
	\hspace{-0.5cm}  
	\begin{subfigure}[t]{0.2\textwidth}
		\includegraphics[width=4cm,angle=0] 
		{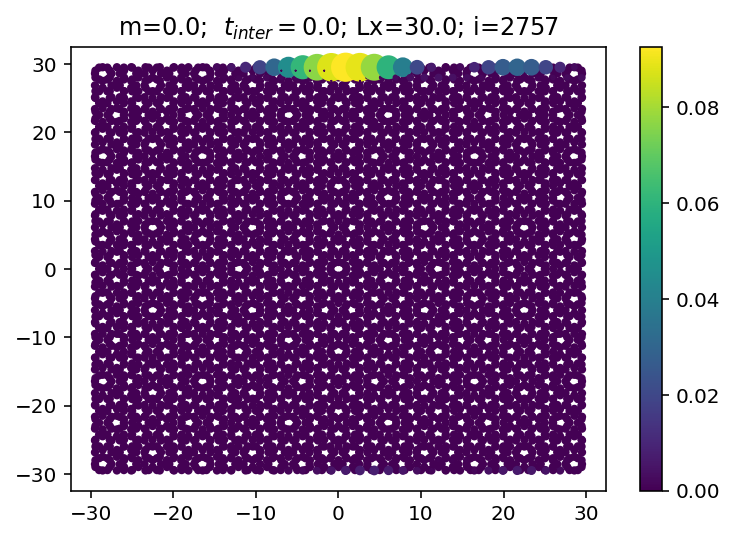}
		\subcaption{$t_{inter}=0.0$, $i=2757$, $E=0.0$}
		\label{Graphene_m_0.0_t_2757}
	\end{subfigure}
	\begin{subfigure}[t]{0.2\textwidth}
		\includegraphics[width=4cm,angle=0] 
		{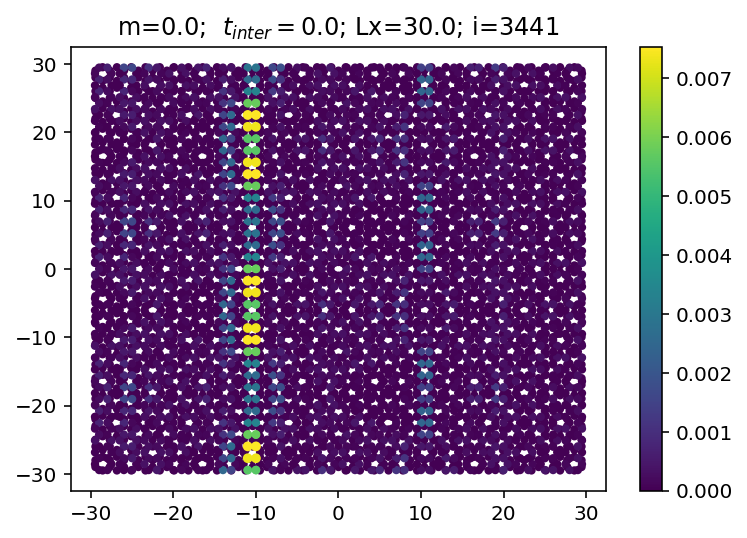}
		\subcaption{$t_{inter}=0.0$, $i=3441$, $E=1.0$}
		\label{Graphene_m_0.0_t_3441}
	\end{subfigure}
	\begin{subfigure}[t]{0.2\textwidth}
		\includegraphics[width=4cm,angle=0] 
		{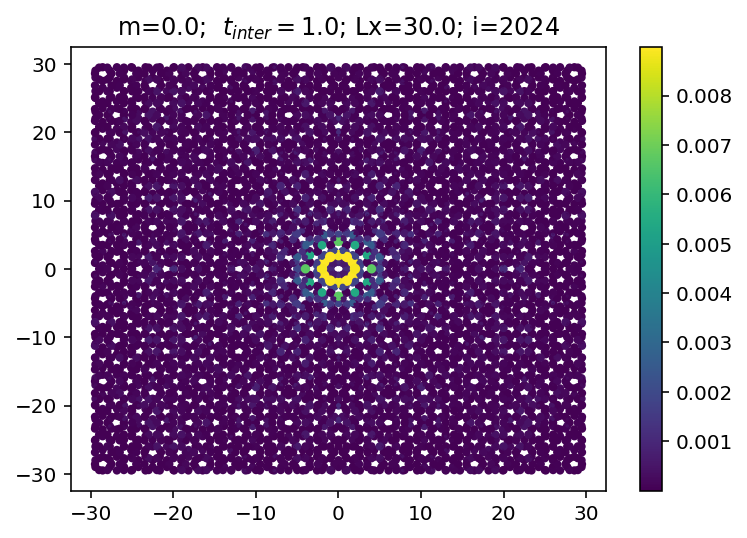}
		\subcaption{$t_{inter}=1.0$, $i=2024$, $E=-1.05$}
		\label{Graphene_m_0.0_t_2024}
	\end{subfigure}
	\begin{subfigure}[t]{0.2\textwidth}
		\includegraphics[width=4cm,angle=0] 
		{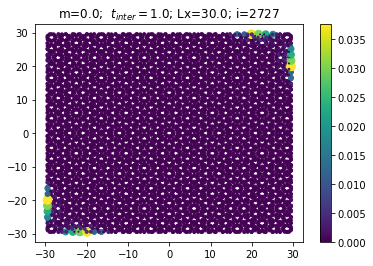}
		\subcaption{$t_{inter}=1.0$, $i=2727$, $E=-0.18$}
		\label{m_0.0_t_2727}
	\end{subfigure}
	\begin{subfigure}[t]{0.2\textwidth}
		\includegraphics[width=4cm,angle=0] 
		{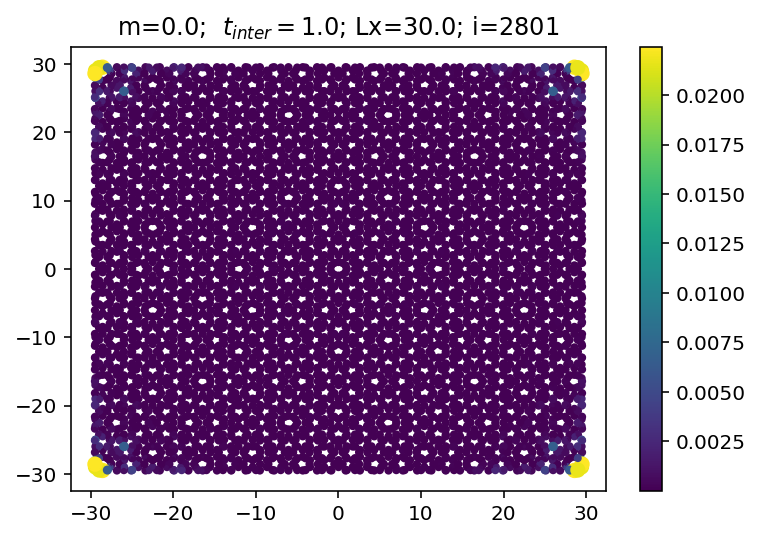}
		\subcaption{$t_{inter}=1.0$, $i=2801$, $E=0.11$}
		\label{Graphene_m_0.0_t_2801}
	\end{subfigure}
	\begin{subfigure}[t]{0.2\textwidth}
		\includegraphics[width=4cm,angle=0] 
		{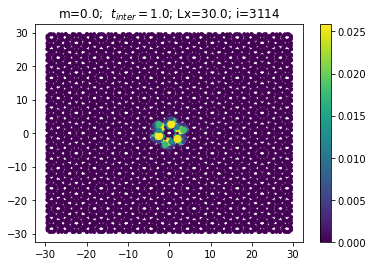}
		\subcaption{$t_{inter}=1.0$, $i=3114$, $E=0.24$}
		\label{m_0.0_t_3114}
	\end{subfigure}
\caption{$|\psi_i|^2$ for a few of the states in $30^{\circ}$ twisted bilayer graphene (we choose $L_x = L_y = 30$, and the other parameters are listed in the text of the Sec. \ref{Sec:twisted_bilayer_graphene}). We observe that at $t_{inter}=0$, the model hosts armchair edge states at $E=0$ (an example is shown on the Fig. \ref{Graphene_m_0.0_t_2757}), as well as localized bulk states at $E= \pm 1$ (an example is shown on the Fig. \ref{Graphene_m_0.0_t_3441}). On the other hand, at $t_{inter}=1$, the system still hosts edge states (\ref{m_0.0_t_2727}), but in addition, it hosts states localized at the corners (\ref{Graphene_m_0.0_t_2801}), as well as at the center of the lattice (\ref{Graphene_m_0.0_t_2024}, \ref{m_0.0_t_3114}). 
}
\label{Localized_states}
\end{figure}


\section{Calculation of the entanglement entropy}
\label{Sec:Calculation_of_the_entanglement_entropy}

Here we briefly describe the method we use to compute entanglement entropy, which was proposed in Refs. \cite{Helmes:2016ahb, Ingo_Peschel_2003} and used in a number of works later on, e.g. in \cite{PhysRevB.99.155153}. We consider the model of twisted bilayer given by the Hamiltonian (\ref{MainHamiltonian}) and are interested in entanglement entropy of any subsystem from the Fig. \ref{Entanglement_figures_cropped}. First, we compute eigenvectors $\psi$ of the Hamiltonian (\ref{MainHamiltonian}) and select only the ones corresponding to its filled states. Next, we compute 'correlation matrix', i.e. sum of their outer products 
\begin{eqnarray}
C_{ij} = \sum\limits_{f - filled \atop states} \psi_f (\vec{r}_i) \psi_f^{\dagger} (\vec{r}_j).
\end{eqnarray}
After that we select the components whose coordinates $\vec{r}_i, \vec{r}_j$ are inside the subsystem whose entanglement entropy we are interested in and thus obtain 'reduced correlation matrix' $C_{A, ij}$. As it was shown in \cite{Helmes:2016ahb, Ingo_Peschel_2003}, the entanglement entropy can be expressed in terms of its eigenvalues $\zeta_m$ as
\begin{eqnarray}
S = - \sum_{m} 
\left(
\zeta_m \log \zeta_m
+ (1 - \zeta_m)  \log (1 - \zeta_m).
\right)
\nonumber\\
\end{eqnarray}
Finally, to substract the leading 'area law' terms and extract the topological contribution, we compute the sum of entanglement entropies over several figures as shown on the Fig. \ref{Entanglement_figures_cropped}.


\section{Kubo formula for anomalous Hall effect}
\label{Sec:KuboFormulaAHE}

In this section, we derive an expression for the anomalous Hall conductivity on a finite lattice using Kubo formula. Importantly, we do not use momentum representation, which in turn makes our derivation applicable to the case of non-periodic lattice. 

We start from partition function of the system, which we write as
\begin{eqnarray}
\mathcal{Z} = \int D \psi D \bar{\psi}
e^{  \sum\limits_{w, i} i w c^{\dagger}_{i, w} c_{i, w} - \sum\limits_{w, i, j} c^{\dagger}_{i, w} H_{ij} c_{j, w} 
} .
\end{eqnarray}
In this expression, we have transformed imaginary time into Matsubara frequencies representation, but left the coordinate representation. Here $w$ denote the Matsubara frequencies and $H_{ij}$ is the lattice Hamiltonian. The indices $i, j$ numerate the lattice sites. Electric field is introduced via Peierls substitution, which in turn may be approximated as
\begin{eqnarray}
H_{ij} \to H_{ij} e^{i \int\limits_{r_i}^{r_j} \vec{A} \vec{dr}}
\approx H_{ij} + i H_{ij} \vec{A} (\vec{r}_j - \vec{r}_i).
\end{eqnarray}
From the last equation we can see that the current components have the form
\begin{eqnarray}
\vec{J}_{ij, \Omega} = i \sum\limits_{w} c^{\dagger}_{i, w+\Omega} H_{ij} (\vec{r}_j - \vec{r}_i) c_{j, w}
\end{eqnarray}
Note, that the current $\vec{J}_{ij, \Omega}$ at Matsubara frequency $\Omega$ is defined here as a vector for each bond connecting the lattice sites $i$, $j$.

Now we apply Kubo formula and write that the $x$ component of the expectation current has a form
\begin{eqnarray}
J_{x, ij, \Omega} = 
\langle 
J_{x, ij, \Omega} \sum\limits_{k. l} J_{y, kl, -\Omega}
\rangle A_{y, kl, \Omega},
\end{eqnarray}
where $ A_{y, kl, \Omega}$ is a $y$ component of the vector potential at the bond connecting the lattice sites $k, l$ at Matsubara frequency $\Omega$.
Now we assume that the electric field is constant in space, and the vector potential is written in the gauge $\vec{A}_{\Omega} = \vec{E}_{\Omega}/(i\Omega)$. The anomalous Hall conductivity is given by the antisymmetric part of the current-current correlator, namely
\begin{eqnarray}
\sigma^{(Hall)}_{ij} = \frac{1}{2i\Omega} 
\sum\limits_{kl}
\left(
\langle J_{x, ij, \Omega} J_{y, kl, -\Omega} \rangle
- \langle J_{y, ij, \Omega} J_{x, kl, -\Omega} \rangle
\right).
\nonumber\\
\label{AHE_Kubo_formula}
\end{eqnarray}
Note that in our notations, the conductivity has just two indices $i, j$, which physically describe electric current at the bond connecting lattice sites $i, j$ as a response to constant electric field (if the electric field was not constant, the conductivity would have four components: two referring to the bond whose electric current we are interested and two referring to the bond whose electric field we look at).

We evaluate the current-current correlator in a conventional way by substituting fermionic Green's functions and summing over Matsubara frequencies. The Green's function can be written in terms of eigenstates $\psi_{n}(\vec{r}_i)$ and energies $E_n$ (here $n$ numerates the states) as 
\begin{eqnarray}
G_{ij, w} = \sum\limits_n \frac{\psi_{n}(\vec{r}_i) \psi^{\dagger}_{n} (\vec{r}_j) }{i w - E_n}.
\end{eqnarray}
After we sum over Matsubara frequencies, assume zero temperature and take the limit of zero external frequency, we can obtain an answer for $\sigma^{(Hall)}_{ij}$, which is given by the Eq. (\ref{AHE_gen_expr}).

Finally, let us look at the sum  $\sum\limits_{ij}\sigma^{(Hall)}_{ij}$ over all lattice sites $i, j$. We can simplify the Eq. (\ref{AHE_gen_expr}) by using explicitly the fact that $\psi_n (\vec{r}_i)$ are eigenvectors, namely by applying an identity
\begin{eqnarray}
\sum\limits_{k, l} \psi^{\dagger}_{f, k} H_{k, l}(x_l - x_k) \psi_{e, l}
= (E_f - E_e) \sum\limits_{k} \psi^{\dagger}_{f, k}  x_k \psi_{e, k}.
\nonumber\\
\end{eqnarray}
We remind that in these notations $\psi_{f, k} \equiv \psi_f (\vec{r}_k)$ and $\psi_{e, k} \equiv \psi_e (\vec{r}_k)$. Also in this equation, $x_k$ means $x$ coordinate of the lattice site $k$, and in the same way $x_l$ means $x$ coordinate of the lattice site $l$.
We can obtain that the total sum has the form
\begin{eqnarray}
&&  \sum\limits_{i, j }  \sigma^{(Hall)}_{ij}
= \frac{1}{2} \sum\limits_{f - filled \atop e - empty} 
\sum\limits_{i, k}
\\
&&  \times
\left\{
\psi^{\dagger}_{e, i} x_i \psi_{f, i} \cdot \psi^{\dagger}_{f, k} y_k \psi_{e, k} 
- \psi^{\dagger}_{f, i} H_{ij} x_i \psi_{e, i}\cdot  \psi^{\dagger}_{e, k} y_k \psi_{f, k} 
\right.  
\nonumber\\
&&  \left. 
- \psi^{\dagger}_{e, i} y_i \psi_{f, i}\cdot  \psi^{\dagger}_{f, k} x_k \psi_{e, k} 
+ 
\psi^{\dagger}_{f, i} y_i \psi_{e, i}\cdot   \psi^{\dagger}_{e, k} x_k \psi_{f, k} 
\right\}.
\nonumber
\end{eqnarray}
Here the summation is taken over filled states numerated by an index $f$ and empty states numerated by an index $e$, as well as lattice sites numerated by indices $k, l$.
After applying the completeness relation
\begin{eqnarray}
\sum\limits_e \psi_{e, i}\cdot  \psi^{\dagger}_{e, k}
= 1 - \sum\limits_f \psi_{f, i}\cdot  \psi^{\dagger}_{f, k}
\label{Completeness_relation}
\end{eqnarray}
one can derive that the sum $\sigma_{ij}$ is equal to 
\begin{eqnarray}
\sum\limits_{i, j }  \sigma^{(Hall)}_{ij}
= \frac{1}{2} \sum\limits_{f} 
\sum\limits_{i}
\left\{
\psi^{\dagger}_{f, i}( y_i x_i  - x_i y_i) \psi_{f, i} 
\right\},
\nonumber\\
\end{eqnarray}
i.e. it is indeed zero for any finite lattice. From the last expression, one can also see how this argument may break down for an infinite lattice: we found that anomalous Hall conductivity is proportional to a trace of a commutator $[X, Y]$. The latter is always zero in a finite-dimensional space, but does not have to be zero in an infinite-dimensional space.


\section{Anomalous Hall conductivity on an infinite crystalline lattice}
\label{Sec:AHE_crystalline}

The goal of this section is to show that the expression for Hall conductivity (\ref{AHE_gen_expr}) from the main text in the case of an infinite crystalline lattice is indeed equivalent to its well-known expression in terms of Berry curvature. Here, for clarity we separate lattice indices $i, j. ...$ into pairs $(i, \alpha), (j, \beta), ...$, which refer to unit cells and sublattice degrees of freedom respectively.  

The key feature of a crystalline lattice is that its eigenvectors are plane waves
\begin{eqnarray}
\psi_{n, i\alpha} = \frac{1}{\sqrt{N}} e^{i \vec{k} \vec{r}_i } \phi_{n, \alpha, \vec{k}}.
\label{Psi_plane_wave}
\end{eqnarray}
Here the wavefunctions $\phi_{n, \alpha, \vec{k}}$ are eigenfunctions of the Schrodinger equation in momentum representation 
\begin{eqnarray}
E_{n, \vec{k}} \phi_{n, \alpha, \vec{k}} = \sum\limits_{\beta} H_{\alpha \beta} (\vec{k}) \phi_{n, \beta, \vec{k}}
\end{eqnarray}
and the corresponding momentum-space Hamiltonian is just a Fourier transformation of the original Hamiltonian
\begin{eqnarray}
H_{\alpha \beta}(\vec{k}) = 
\sum\limits_{j, \beta}
e^{i \vec{k}(\vec{r}_j - \vec{r}_i) }
 H_{i,j \atop \alpha, \beta}. 
\end{eqnarray}
Due to translational symmetry, we can reverse the above expression and thus write it as 
\begin{eqnarray}
H_{i,j \atop \alpha, \beta} 
= \frac{1}{N} \sum\limits_{\vec{k}} H_{\alpha\beta}(\vec{k}) e^{-i \vec{k} (\vec{r}_j - \vec{r}_i) }.
\end{eqnarray}
After substituting the above expression, as well as expressions for the eigenstates in the form of plane waves (\ref{Psi_plane_wave}) into the main expression for anomalous Hall conductivity (\ref{AHE_gen_expr}), one can obtain the lattice version of TKNN formula
\begin{eqnarray}
\sum\limits_{j} \sigma^{(Hall)}_{i,j \atop \alpha, \beta}
&=& \frac{1}{2} \sum\limits_{\gamma, \delta}
\sum\limits_{f - filled \atop e - empty}
 \frac{1}{N} \sum\limits_{\vec{k}}
\frac{1}{(E_{f, \vec{k}} - E_{e, \vec{k}})^2}
\label{Lattice_TKNN}
\\
&&\times 
\left\{
\phi^{\dagger}_{e, \alpha, \vec{k}} \frac{\partial H_{\alpha\beta}(\vec{k})}{\partial k_x} \phi_{f, \beta, \vec{k}}
\cdot
\phi^{\dagger}_{f, \gamma, \vec{k}} \frac{\partial H_{\gamma\delta} (\vec{k})}{\partial k_y} \phi_{e, \delta, \vec{k}}
\right.
\nonumber\\
&&\qquad -
\phi^{\dagger}_{f, \alpha, \vec{k}} \frac{\partial H_{\alpha\beta}(\vec{k})}{\partial k_x} \phi_{e, \beta, \vec{k}}
\cdot
\phi^{\dagger}_{e, \gamma, \vec{k}} \frac{\partial H_{\gamma\delta} (\vec{k})}{\partial k_y} \phi_{f, \delta, \vec{k}}
\nonumber\\
&& 
\hspace{3cm}
\left.
\phantom{ \frac{\partial H_{\gamma\delta} (k)}{\partial k_y} }
- \left( k_x \leftrightarrow k_y \right)
\right\}.
\nonumber
\end{eqnarray}
Notice that in the left side of this equation we have sum $\sum\limits_j \sigma_{ij}$ of conductivities through all bonds adjacent to a site $i$ - the quantity we discuss in the main text. In addition, to obtain the physical answer we have to sum over all sublattice degrees of freedom, i.e. over the indices $\alpha, \beta$. We also note that in the Eq. (\ref{Lattice_TKNN}) we still assume that the lattice is discrete, but the momentum is continuum because the lattice is infinite. Hence it is straightforward to replace the momentum summation with an integration 
\begin{eqnarray}
  \frac{1}{N} \sum\limits_{\vec{k}} \to S \int \frac{d^2 k}{(2\pi)^2},
\end{eqnarray}
 where $S$ is an area of the unit cell. 

Finally, by making use of identities
\begin{eqnarray}
 \sum\limits_{\alpha \beta}
\phi^{\dagger}_{e, \alpha, \vec{k}} \frac{\partial H_{\alpha \beta }(\vec{k})}{\partial k_x} \phi_{f, \beta, \vec{k}}
= 
\sum\limits_{\alpha} (E_{f, \vec{k}} - E_{e, \vec{k}} ) \phi^{\dagger}_{e, \alpha, \vec{k}} \frac{\partial \phi_{f, \alpha, \vec{k}} }{\partial k_x}
\nonumber\\
\end{eqnarray}
combined with the completeness relation (\ref{Completeness_relation}) we obtain a familiar expression for anomalous Hall conductivity in terms of Berry curvature
 \begin{eqnarray}
 \sum\limits_{j \atop \alpha\beta} \sigma_{i, j \atop \alpha, \beta}
 = S \int \frac{d^2 k}{(2\pi)^2} \sum\limits_{\alpha}
 \left\{
 \frac{\partial \phi^{\dagger}_{f, \alpha} }{\partial k_y}
 \frac{\partial \phi_{f, \alpha} }{\partial k_x}
 -
 \frac{\partial \phi^{\dagger}_{f, \alpha} }{\partial k_x}
 \frac{\partial \phi_{f, \alpha} }{\partial k_y}
 \right\}.
\nonumber\\
 \end{eqnarray}



\newpage

\bibliographystyle{apsrev4-2}
\bibliography{Quascrystals_refs}

\end{document}